\documentclass[11p]{scrartcl}
 \usepackage[top=2.5cm, bottom=2.5cm, left=2.5cm, right=2.5cm]{geometry}

\usepackage[T1]{fontenc}
\usepackage[utf8]{inputenc}
\usepackage{microtype}
\usepackage[english]{babel}
\usepackage{tikz}

\usepackage{pdfpages}

\usepackage{tikz}
\usepackage{pgfplots}
\pgfplotsset{compat=1.5}

\usepackage{amsmath}
\usepackage{amsfonts}
\usepackage{amssymb}
\usepackage{mathrsfs}
\usepackage{amstext}
\usepackage{bbold}
\usepackage{mathtools}
\usepackage{physics}
\usepackage{nicefrac}
\usepackage{dsfont}
\usepackage{cleveref}

\newcommand{\be}{\begin{equation}}

\newcommand{\ee}{\end{equation}}

\newcommand{\ba}{\begin{eqnarray}}
\newcommand{\ea}{\end{eqnarray}}

\usepackage{graphicx}
\usepackage{wrapfig}
\usepackage{multicol}
\usepackage[e]{esvect}
\usepackage[per-mode=reciprocal, exponent-product=\cdot, separate-uncertainty=true, emulate=units]{siunitx}
\usepackage{gensymb}
\usepackage{amsthm}
\usepackage{subfig}
\usepackage{caption}
\usepackage{xcolor}
\usepackage{placeins}

\newcommand*{\myZ}{\mathcal{Z}}

\newcommand*{\jz}{\oper{j}_z}
\newcommand*{\kp}{\oper{k}_+}
\newcommand*{\kx}{\oper{k}_x}
\newcommand*{\ky}{\oper{k}_y}
\newcommand{\km}{\oper{k}_-}
\newcommand*{\kpm}{\oper{k}_\pm}

\newcommand*{\JZ}{\oper{J}_z}
\newcommand*{\KP}{\oper{K}_+}
\newcommand{\KM}{\oper{K}_-}

\newcommand*{\ogHam}{\oper{H}_g}

\newcommand*{\oa}{\oper{a}}
\newcommand*{\oad}{\oper{a}^{\dagger}}

\usepackage{array}
\usepackage{tabularx}
\usepackage{tabulary}
\usepackage{booktabs}
\usepackage{longtable}

\newcommand{\IN}{\mathds{N}}

\newcommand{\IR}{\mathds{R}}
\newcommand*{\IC}{\mathds{C}}

\newcommand*{\e}[1]{e^{#1}}

\newcommand*{\einhalb}{\frac{1}{2}}
\newcommand*{\al}[1]{\mathfrak{#1}}
\newcommand*{\gr}[1]{#1}

\newcommand*{\id}{\mathds{1}}

\newcommand*{\oper}[1]{\hat{#1}}

\newcolumntype{C}{>{$}c<{$}}
\newcolumntype{L}{>{$}l<{$}}
\newcolumntype{R}{>{$}r<{$}}

\newcommand*{\DD}[2][]{\ensuremath{\mathcal{D}^{#1}#2}\,}

\newcommand*{\punkt}{\text{.}}
\newcommand*{\komma}{\text{,}}
\newcommand*{\cas}{\mathfrak{C}}

\newcommand*{\fac}{!\,}

\newcommand{\osin}{\widehat{\sin{(\lambda b)}}}
\newcommand{\BR}[1]{\left( #1 \right)}

\usepackage[normalem]{ulem}

\title{\textsf{{Path integral renormalisation in loop quantum cosmology}}\vspace{0.35cm}}

\author{
{Norbert Bodendorfer${}^1$\footnote{\texttt{norbert.bodendorfer@physik.uni-r.de}}, Muxin Han${}^{2,3}$\footnote{\texttt{hanm@fau.edu}}, Fabian Haneder${}^1$\footnote{\texttt{fabian.haneder@physik.uni-r.de}}}, Hongguang Liu${}^3$\footnote{\texttt{hongguang.liu@gravity.fau.de}}\\
{{${}^1$ Institute for Theoretical Physics, University of Regensburg,}}\\
{{93040 Regensburg, Germany}}\\
{{${}^2$ Department of Physics, Florida Atlantic University,}}\\
{{777 Glades Road, Boca Raton, FL 33431-0991, USA}}\\
{{${}^3$ Institut f\"ur Quantengravitation, Universit\"at Erlangen-N\"urnberg,}}\\
{{Staudtstr. 7/B2, 91058 Erlangen, Germany}}\\
}

\renewcommand{\bf}{\bfseries}
\renewcommand{\tt}{\ttfamily}

\begin{document}

\maketitle

\begin{abstract}
{A coarse graining technique akin to block spin transformations that groups together fiducial cells in a homogeneous and isotropic universe has been recently developed in the context of loop quantum cosmology. The key technical ingredient was an SU$(1,1)$ group and Lie algebra structure of the physical observables as well as the use of Perelomov coherent states for SU$(1,1)$. It was shown that the coarse graining operation is completely captured by changing group representations. Based on this result, it was subsequently shown that one can extract an explicit renormalisation group flow of the loop quantum cosmology Hamiltonian operator in a simple model with dust-clock. 

In this paper, we continue this line of investigation and derive a coherent state path integral formulation of this quantum theory and extract an explicit expression for the renormalisation-scale dependent classical Hamiltonian entering the path integral for a coarse grained description at that scale. We find corrections to the non-renormalised Hamiltonian that are qualitatively similar to those previously investigated via canonical quantisation. In particular, they are again most sensitive to small quantum numbers, showing that the large quantum number (spin) description captured by so called ``effective equations'' in loop quantum cosmology does not reproduce the physics of many small quantum numbers (spins). Our results have direct impact on path integral quantisation in loop quantum gravity, showing that the usually taken large spin limit should be expected not to capture (without renormalisation, as mostly done) the physics of many small spins that is usually assumed to be present in physically reasonable quantum states. 
 }
\end{abstract}

\section{Introduction}

Loop quantum gravity \cite{ThiemannModernCanonicalQuantum, RovelliBook2} is a non-perturbative approach to quantum gravity that ultilises a formulation of general relativity in terms of connection variables as a classical starting point. 
This choice of variables introduces an additional local gauge invariance under the action of a Lie group, akin to Yang-Mills theory. 
In the standard formulation of loop quantum gravity in $3+1$ spacetime dimensions using Ashtekar-Barbero variables \cite{BarberoRealAshtekarVariables}, this group is SU$(2)$.
As a consequence, one encounters a discrete notion of quantum geometry with eigenvalues of geometric operators determined by group invariants. 

Since the famous result of Ponzano and Regge \cite{PonzanoSemiclassicalLimitOf}, it was understood that this type of theory simplifies dramatically in the limit of large quantum numbers, e.g. large spins in the above $3+1$-dimensional theory. More specifically, one obtains a discrete version of general relativity as the leading order in the large quantum number expansion, where the discreteness scale is much larger than the Planck scale. This limit has been termed ``semiclassical'' in the literature and is the best investigated limit in loop quantum gravity, both in the canonical as well as in the path integral (spin foam) formulations. Similarly, it is also the best investigated limit in loop quantum cosmology \cite{SinghLoopQuantumCosmologyABrief}, which is well described by so-called ``effective equations'' in this limit. Here, large quantum numbers translate to a large volume of the universe throughout the evolution. 

There are however several reasons to doubt that the ``semiclassical'' limit of loop quantum gravity is the limit describing our observed universe. Let us mention two of them. First, for a given geometry, say of a certain total volume, there are many more ways to construct such a geometry from smaller building blocks than there are from larger building blocks. Hence, from a maximisation of entropy standpoint, we expect small quantum numbers to dominate. Computations of black hole entropy via state counting \cite{AshtekarQuantumGeometryAndBlackHoleEntropy, DomagalaBlackHoleEntropy}, as well as entanglement entropy \cite{HanSemiclassicalBehaviorOf}, are consistent with this expectation. Second, the large quantum number limit has a natural explanation in terms of ``transplanckian'' high-energy degrees of freedom akin to black holes, that are transplanckian in the sense of energies above the Planck energy, but not in the sense of distances below the Planck length \cite{BNI}. The scaling of the involved actions with the geometric quantities in the large quantum number limit is consistent only with this high-energy interpretation, but not with one where large quantum numbers refer to coarse grained flat space. 

It should be noted that a large quantum number description of our universe may still arise from a coarse graining procedure, where many small quantum numbers are grouped together to large ones, akin to block-spin transformations. Then however, one would expect that the corresponding Hamiltonian operator or path integral kernel transform under a renormalisation group flow that ensures that the coarse dynamics agrees with the fundamental one. In contrast, in the ``semiclassical'' limit mentioned above, one uses the fundamental small quantum number Hamiltonian even for large quantum numbers, which is equivalent to claiming that the Hamiltonian operator is invariant under the renormalisation group flow.

Addressing coarse graining directly is often impossible in analytical calculations. Hence, despite an increased amount of attention devoted to the topic \cite{MarkopoulouCoarseGrainingIn, OecklRENORMALIZATIONFORSPIN, LivineCouplingOfSpacetime, DittrichCoarseGrainingMethods, BahrHolonomySpinFoam, BahrOnBackgroundIndependent, BahrHypercuboidalRenormalizationIn, BahrNumericalEvidenceFor, CarrozzaFlowingInGroup, BahrRenormalizationOfSymmetry, DittrichCoarseGrainingFlow, BodendorferStateRefinementsAnd, BodendorferCoarseGrainingAs, ThiemannRenormalisationReview}, concrete results connecting small with large quantum numbers are scarce. Recently however, progress was made in a simplified cosmological model, relying heavily on an SU$(1,1)$ Lie group structure \cite{BojowaldDynamicalCoherentStates, BojraDynamicsForA, LivineGroupTheoreticalQuantization, BenAchourThiemannComplexifierIn} that allowed to analytically compute a coarse graining operation from small to large spins \cite{BodendorferCoarseGrainingAs, BWI, BHAddendum}. In particular, the renormalisation group flow of the Hamiltonian operator, which shares key features with similar operators in full loop quantum gravity, was explicitly computed \cite{BWI}. It was found that this flow is non-trivial and introduces significant corrections to the renormalised operator. Moreover, it was shown that neglecting to account for this renormalisation group flow changes the physics of the model. While the results were obtained in a simple model, we expect the main result, that Hamiltonian operators of the type used in loop quantum gravity have a non-trivial renormalisation group flow, applies also to more complicated models and eventually to full loop quantum gravity. In fact, the model used here can be understood as arising from a full quantum gravity theory of loop quantum gravity type via a suitable truncation \cite{BIII, BVI, BodendorferStateRefinementsAnd}.

In this paper, we will generalise the results of \cite{BodendorferCoarseGrainingAs, BWI, BHAddendum}, which were so far restricted to canonical quantisations, to path integrals. To this end, we rewrite the quantum theory of \cite{BodendorferCoarseGrainingAs} as a coherent state path integral, making use of the Perelomov coherent states that are crucially involved in the coarse graining map. We further explain how the coarse graining operation studied in \cite{BodendorferCoarseGrainingAs, BHAddendum} translates to path integrals. Finally, we rewrite the coherent state path integral as a phase space path integral where the path integration is over the standard loop quantum cosmology variables $v$, $b$. From this expression, we can also extract the renormalised (classical) Hamiltonian that enters the path integral measure through the canonical action of the standard form $S = \int \dd{t} (p \dot q -H)$.

The coarse graining operation in this work closely resembles a similar operation in the context of spin foam models \cite{HanSemiclassicalBehaviorOf}, where the states in the large spin spin foam model are tensor products of states with small spins, due to the same factorization property of coherent states that enters the coarse graining map in this paper \cite{BHAddendum}. The large spin model is emergent as an effective theory, via a block-spin-like coarse graining procedure, from a more fundamental spin foam model with small spins and more refined graphs. This description, as opposed to the naive ``semiclassical'' large spin limit involving non-divisible large chunks of geometry, has the advantage of obtaining the anticipated area law when computing entanglement entropy. Despite this success, it was not possible to obtain the coarse grained dynamics in \cite{HanSemiclassicalBehaviorOf} due to the complexity of the full $4d$ spin foam model. Hence, to gain some insight into how path integral kernels of the type encountered in spin foam models could renormalise, it is necessary to study simplified models. 

For this, the present paper is an example. 
The path integral studied here is a loop quantum cosmology analogue of spin foam models (see e.g. \cite{AshtekarCastingLoopQuantum} for earlier results on the spin foam formulation of Loop Quantum Cosmology as well as \cite{HanEffectiveDynamicsFrom} for a coherent state path integral formulation of full Loop Quantum Gravity), but can also be considered as a truncation of full quantum gravity theories subject to suitable homogeneity assumptions \cite{BIII, BVI, BodendorferStateRefinementsAnd}.
We expect that the key lesson drawn from this paper, as well as \cite{BWI}, i.e. that the renormalisation group flow of the Hamiltonian is non-trivial and affects physics, also applies to full models, both in the canonical and path integral formulations. 

This paper is organised as follows:\\
Section \ref{sec:GI} provides an overview of \cite{BodendorferCoarseGrainingAs, BWI, BHAddendum}, on which the present work is based. In particular, section \ref{sec:GroupQuantisation} defines the quantum system that we later formulate as both a coherent state and phase space path integral. Section \ref{sec:CSPIrecap} provides a review of coherent state path integrals at the example of the harmonic oscillator and the Bose-Hubbard model, and discusses an important pitfall that occurs in such formulations. The main results of the paper are derived in section \ref{sec:CSPIsu11}, including the coherent state path integral with continuum action \eqref{eq:ActionContinuumSpinor}, as well as the phase space path integral with renormalised Hamiltonian \eqref{eq:HPIvb}. Some implications of our work are discussed in section \ref{sec:Discussion}. Appendix \ref{app:SU11} provides an overview of the representation theory of su$(1,1)$ relevant for this work. Appendix \ref{sec:identity} discusses the resolution of the identity with particular care about the special case $j=\nicefrac{1}{2}$.

\section{Group quantisation, coarse graining map, and renormalised operators} \label{sec:GI}

In this section, we briefly review the group quantisation strategy to arrive at a quantum cosmology model and its specific application to loop quantum cosmology. We also review how group quantisation naturally leads to a notion of coarse graining under the assumption of homogeneity and non-interaction of neighbouring spatial points, as is usually assumed in cosmology. A more detailed review is given in \cite{BWI}. 

\subsection{Classical description, polymerisation, and group quantisation} \label{sec:GroupQuantisation}

\subsubsection{Classical formulation}

Classically, the gravitational sector of spatially flat, homogeneous and isotropic cosmology can be described by the canonical pair $\{b,v\}=1$, where $v$ is the volume of the spatial slice (either compact or regularised by using a fiducial cell), and $b=-3\frac{\dot a}{a}$ is proportional to the Hubble rate. In units where $12 \pi G = \hbar = c = 1$, the Hamiltonian constraint reads
\be
	\mathcal H = \mathcal H_m + \mathcal H_g, ~~~~~ \mathcal H_g = - \frac{v b^2}{2} \label{eq:Hclass}
\ee
where $\mathcal H_m$ and $\mathcal H_g$ denote the matter and gravitational parts respectively. As is well known, the vacuum ($\mathcal H_m = 0$) solution of this model is given by Minkowski space, so that a matter contribution to the Hamiltonian is essential in obtaining non-trivial dynamics. 

In this paper, we are going to use non-rotational dust as our matter field \cite{BrownDustAsStandard, SwiezewskiOnTheProperties}, mainly for simplicity. Making this choice, one can deparametrise the theory by using the value of the dust field as a clock and obtains a true Hamiltonian $H$ that generates time evolution in the dust time, given simply by $H=\mathcal H_g$. 

It is well known that the evolution generated by \eqref{eq:Hclass} leads to a singularity either in the future (``big crunch'') or in the past (``big bang''), whose desired resolution has been the main motivation for constructing a quantum theory of cosmology. One such candidate theory is loop quantum cosmology, see \cite{AshtekarLoopQuantumCosmology, SinghLoopQuantumCosmologyABrief} for reviews. A key feature of loop quantum cosmology is that an operator corresponding to $b$ does not exist, while the exponentials $e^{i \lambda b}$ with $\lambda$ being a free parameter at the Planck scale exist. This peculiarity leads to Hamiltonians where $b$ has to be substituted by a sum of exponentials. The most adopted choice is to substitute
\be	
	b \mapsto \frac{\sin( \lambda b)}{\lambda}\text{,}
\ee
so that the resulting expression reduces to $b$ in the low curvature limit $b \ll \lambda^{-1}$, but other expressions are possible as well. Applied to $\mathcal H_g$, this leads to 
\be
	\mathcal H_g \mapsto - \frac{v }{2} \frac{ \sin^2(\lambda b)^2}{\lambda^2} \text{.} \label{eq:HHol}
\ee
In particular, one can do this substitution already at the classical level, where it is often called ``polymerisation''. It turns out that the so-obtained classical theories give good approximations to the results of the quantum theory if the quantum numbers remain large throughout the evolution. In this case, one refers to the classical equations of motion as ``effective equations''.  

While so-called holonomy corrections as in \eqref{eq:HHol} are forced upon us from the Hilbert space construction, also corrections for small volumes, i.e. of about the Planck volume, can reasonably be expected from a quantum theory and appear in explicit constructions of the Hamiltonian constraint in full loop quantum gravity \cite{ThiemannQSD1}, as well as loop quantum cosmology \cite{AshtekarQuantumNatureOf}. In particular, we will later find that the phase space path integral is determined by the Hamiltonian
\be
	H_{\text{PI}} = -\frac{1}{2}\frac{\sin[2](\lambda b)}{\lambda^2}\sqrt{v^2-v_m^2}	+\frac{1}{4\lambda^2}\left(\sqrt{v^2-v_m^2}-v\right), \label{eq:HClassVm}
\ee
which introduces explicitly a minimal volume $v_m$ and reduces to \eqref{eq:HHol} in the limit $v \gg v_m$. 

For the purpose of group quantisation, we specifically consider the three phase space functions
\be
	j_z = \frac{v}{2\lambda}, ~~~ k_\pm = \frac{\sqrt{v^2-\sigma}}{2 \lambda} e^{\pm 2 i \lambda b},  \label{eq:ClassicalGenerators}
\ee
whose Poisson algebra 
\be
	\{j_z, k_\pm \} = \mp i k_\pm, ~~~~ \{ k_p, k_- \} = 2 i j_z \label{eq:ClassCommSU11}
\ee
is isomorphic to the Lie algebra su$(1,1)$, see \eqref{eq:su11commutators}. Here, $\sigma \in \IR$ is a free constant to be matched in the quantum theory to the group representation. In fact, the generators satisfy the constraint
\be
     j_z^2-k_x^2-k_y^2=\sigma/4\lambda^2\text{,}
\ee
where the left hand side can be identified as the classical version of the su$(1,1)$ quadratic Casimir operator that determines the representation. 
For $v^2 \gg \sigma$,
\be
	H = \frac{1}{2 \lambda} \left( \frac{k_+ + k_-}{2}- j_z\right) \label{eq:GeneratorsH}
\ee
reduces to \eqref{eq:HHol}, and we will use the expression \eqref{eq:GeneratorsH} as a Hamiltonian for any value of $\sigma$ for purposes of a straight forward coarse graining (see below). 
Alternatively, we can use the functions $k_x$ and $k_y$ related via 
\be
	k_\pm = k_x \pm i k_y,
\ee 
in which case we have $H = \frac{1}{2 \lambda} \left(k_x - j_z\right)$.

\subsubsection{Quantum theory}

Quantising the Poisson algebra \eqref{eq:ClassCommSU11} directly is straightforward, as it amounts to simply looking up the representation theory of the Lie algebra su$(1,1)$, see appendix \ref{app:SU11}. For our purposes, it is sufficient to consider the discrete series with positive magnetic quantum numbers, where irreducible representations are labelled by a positive half-integer $j \in \mathbb N / 2$. A basis of the representation space for a given $j$ is given by $\ket{j, m}$, $m \in \{j, j+1, j+2, \ldots\}$, and the action of the generators is given by
\ba
		\jz\ket{j,m}&=&m\ket{j,m}\komma\\
		\kp\ket{j,m}&=&\sqrt{m(m+1)-j(j-1)}\ket{j,m+1}\\
		\km\ket{j,m}&=&\sqrt{m(m-1)-j(j-1)}\ket{j,m-1}.
\ea
Via $\hat k_\pm = \hat k_x \pm i \hat k_y$, we may also work out the action of $\kx$ and $\ky$. The quadratic Casimir operator reads 
\be
	\mathfrak C = \jz^2-\kx^2-\ky^2
\ee 
and evaluates to $j(j-1)$ in the above discrete series of representations. 

Let us come back to the constant $\sigma$ in \eqref{eq:ClassicalGenerators} and match it to the representation label $j$. This can be done in several ways, which reflects ambiguities in the definition of the quantum theory. We will discuss three possible choices here and find later that only choice 1 is consistent with the most straight forward definition of the phase space path integral. 
\begin{enumerate}
	\item Since the smallest eigenvalue of the volume operator $\hat v := 2 \lambda \jz$ is given by $2 \lambda j$, we could choose $\sigma=   \left(2 \lambda j\right)^2 =: v_m^2$ as the ``minimal volume''. In this case, the coarse graining map below is well motivated even at intermediate levels of coarse graining as the functions \eqref{eq:ClassicalGenerators} are always extensive in the system volume. Due to this, the choice $\sigma = v_m^2$ was adopted in \cite{BodendorferCoarseGrainingAs} and proposed earlier in \cite{LivineGroupTheoreticalQuantization} for other reasons. With this choice, the classical expression $j_z^2-k_x^2-k_y^2=\sigma/4\lambda^2$ of the Casimir operator evaluates to $(v_m/2\lambda)^2$ and matches the exact quantum expression $j(j-1)$ to leading order in $j$. It should be noted that this choice already introduces a volume gap classically even in the $j=\nicefrac{1}{2}$ representation, since the expressions for $k_\pm$ are well defined only for $v\geq v_m$, and results in the Hamiltonian \eqref{eq:HClassVm} via \eqref{eq:GeneratorsH}.  
	\item One may choose to match the Casimir operator exactly with the classical expression, leading to $\sigma = j(j-1)$. In this case, one loses the interpretation of $\sqrt{\sigma}$ as the minimal volume.
	\item In order to obtain a quantum theory that agrees with the usual constructions of loop quantum cosmology in the case $j=\nicefrac{1}{2}$, we can make a choice so that $\sigma=0$ for $j=\nicefrac{1}{2}$. A simple possibility suggested by the analysis of \cite{BWI} is given by $\sigma=(v_m-\lambda)^2$. In this case, the classical expression for the Casimir evaluates to $\frac{v_m}{2 \lambda}(\frac{v_m}{2 \lambda}-1)+\frac{1}{4}$ and matches $j(j-1)$ to leading and next to leading order. 
\end{enumerate}

\subsubsection{Coherent states}

On the representation spaces, we can introduce the coherent states (see e.g. \cite{BenAchourThiemannComplexifierIn})
	\begin{align}\label{eq:cohStates}
		\ket{j,z}=(2L(z))^j\sum_{m=j}^{\infty}
		\sqrt{\frac{\Gamma(m+j)}{\Gamma(m-j+1)\Gamma(2j)}}
		\frac{(z_1)^{m-j}}{(\bar z_0)^{m+j}}\ket{j,m}\komma
	\end{align}
where \( 2L(z)=\abs{z_0}^2-\abs{z_1}^2 \) and \( z=\smqty(z_0\\\bar z_1)\in\IC^2 \) such that \( \abs{z_0}>\abs{z_1} \). This last requirement is needed so that the states are normalisable. Note the complex conjugation of the second entry in $z$, which will hereinafter be called a \emph{spinor}. These states are constructed in a very similar manner to the well-known coherent states of Perelomov \cite{PerelomovCoherentStatesFor}, but they differ in that there are two complex parameters, rather than only one. This introduces an additional phase symmetry reflecting the redundancy in the description. One can re-obtain standard Perelomov states by fixing \( z_0=1 \). 

One of the most useful properties of these states is how they behave under the action of SU$(1,1)$ operators, given by
\begin{equation}\label{eq:groupOnCohStates}
	\oper{U}\ket{j,z}=\ket{j,Uz},\quad\forall \,U\in\text{SU(1,1)}\komma 
\end{equation}
where $U$ is in the $j$-irrep on the left, and in the defining representation, given by the basis \eqref{eq:definingRep}, on the right. Consequently, the only thing required to calculate the evolution of the coherent states under the SU(1,1) group flow is \( 2\times2 \) matrix multiplication. Given that we will choose the Hamiltonian of the system to be the linear combination of generators in \eqref{eq:GeneratorsH}, we know that the time evolution operator \( \e{-i H t}\in\text{SU(1,1)} \), so that the entire time evolution of the coherent states can be computed by this very simple procedure \cite{LivineGroupTheoreticalQuantization, BWI}, leading to 
\begin{equation}\label{eq:zOfT}
	\pmqty{z_0(t)\\\bar{z}_1(t)}=\exp(-it\ogHam)\pmqty{z_0(0)\\\bar{z}_1(0)}
	=\pmqty{z_0(0)+\frac{it}{4\lambda}(z_0(0)-\bar{z}_1(0))\\
		\bar{z}_1(0)+\frac{it}{4\lambda}(z_0(0)-\bar{z}_1(0))} \text{.}
\end{equation}

As is usually the case \cite{PerelomovBook}, this set of coherent states is overcomplete. Indeed, it admits a resolution of the identity \cite{LivineGroupTheoreticalQuantization},
\begin{equation}\label{eq:identityResolution}
	\id=\int\dd[2]{z_0}\dd[2]{z_1}\frac{(2j-1)}{2\ell\pi^2}\delta(2L(z)-2\ell)\dyad{j,z}\punkt
\end{equation}
One can immediately see that the case \( j=\nicefrac{1}{2} \) is tricky, and we will discuss it in appendix \ref{sec:identity}.

We can also calculate matrix elements of the generators and their products rather easily. The simplest examples are the expectation values of the generators, for which direct computation yields
\ba 
		\ev{\jz}{j,z}&=&j\frac{\abs{z_0}^2+\abs{z_1}^2}{\abs{z_0}^2-\abs{z_1}^2} \label{eq:Ejz}\\
		\ev{\kp}{j,z}&=&j\frac{\bar{z}_0\bar{z}_1}{L} \label{eq:Ekp} \\
		\ev{\km}{j,z}&=&j\frac{z_0z_1}{L}\punkt \label{eq:Ekm}
\ea
The general formula for $p,q,r \in \mathbb N_0$ reads
	\begin{align}
		& \bra{j,\zeta}{\km^p\, \jz^q\, \kp^r}\ket{j,z}	\label{matrixElements} \\
			=&(2L(\zeta))^j(2L(z))^jz^{p-r}\sum_{\mu=0}^{\infty}\frac{\Gamma(\mu+p+2j)\Gamma(\mu+p+1)}{\mu\fac\Gamma(\mu+p-r+1)\Gamma(2j)}\left(\mu+p+j\right)^q\left(\bar{\zeta} z\right)^\mu. \nonumber 
	\end{align}

\subsection{Coarse graining} \label{sec:CG}

Let us now take a closer look at \eqref{eq:Ejz}-\eqref{eq:Ekm}. We observe that the representation label $j$ enters the expectation values simply as a direct proportionality. Since all three generators ($\sigma = v_m^2$) have a classical interpretation of an extensive quantity, i.e. scaling with the system volume, this suggests the following coarse graining scheme \cite{BodendorferCoarseGrainingAs, BHAddendum}:
\begin{itemize}
	\item We consider the quantum system in representation $j_0 ~ \Leftrightarrow ~ v_m = 2 \lambda j_0$ as the true fundamental quantum system. The basic observables of this system are polynomials in the generators $\jz, \kp, \km$. The minimal eigenvalue of the volume operator is $v_m$. We consider the system in a Perelomov coherent state $\ket{j_0, z}$. $z$ encodes the intensive properties of the system, i.e. ratios of extensive quantities. 
	\item It makes most sense to consider $j_0 = \nicefrac{1}{2}$ as the fundamental system that cannot be refined further, because this choice does not put any lower cut-off to the allowed volume eigenvalues. However, for the coarse graining prescription below, this is not necessary. In the case $j_0>\nicefrac{1}{2}$, such a system has already been coarse grained to coarseness $j_0$, i.e. minimally resolved volume $2 \lambda j_0$, and will subsequently be further coarse grained. 
	\item As a coarse grained system, we consider a compound system made up of $N$ non-interacting fundamental systems with identical quantum states, i.e. 
		\be
			\underbrace{\ket{j_0,z}\otimes\ket{j_0,z}\otimes\ldots}_{N\text{~times}}.
			\label{eq:ProductState}
		\ee 
		Due to the extensive nature of the observables, operators in the coarse grained and fundamental descriptions are linked as (and similarly for the other generators $\jz, \kp$)
		\begin{equation}
			\KM\coloneqq\left(\prescript{}{1}{\km}+\prescript{}{2}{\km}+\ldots+\prescript{}{N}{\km}\right), \label{eq:DefCoarseKM} 
		\end{equation}
		where $\prescript{}{i}\km$ is a generator in representation $j$ associated to the quantum system denoted by its pre-script $i$.
	\item As a consequence of the construction of Perelomov coherent states and the associated group theory \cite{BodendorferCoarseGrainingAs, BHAddendum}, it turns out that the coarse grained system is described by a Perelomov coherent state in representation $j = Nj_0$ with spinor label $z$ and the coarse grained operators are simply given by the generators in representation $j = Nj_0$. This leads to the following coarse graining map \cite{BodendorferCoarseGrainingAs, BHAddendum}:
	\begin{center}
\def\arraystretch{2.4}
  \begin{tabular}{ l | c | c }
    
     & Fine description & Coarse description \\ \hline
    Quantum state & $\underbrace{\ket{j_0,z}\otimes\ket{j_0,z}\otimes\ldots}_{N\text{~times}} $ & $\ket{Nj_0, z}$ \\ [20pt] \hline
    Operators & $ \left(\sum_{a=1}^N \prescript{}{a}\km\right)^p \left(\sum_{b=1}^N \prescript{}{b}\jz \right)^q  \left(\sum_{c=1}^N \prescript{}{c}\kp\right)^r$ & $\KM^p \JZ^q \KP^r$ \\ [7pt]
    \hline
  \end{tabular}
\end{center}	
	It holds for arbitrary coherent state matrix elements of polynomials in the generators, i.e.
	\begin{eqnarray}
	\bra{j_0\otimes j_0\otimes\ldots,\zeta}{\KM^p\, \JZ^q\, \KP^r}\ket{j_0\otimes j_0\otimes\ldots,z} = \mel{Nj_0,\zeta}{\KM^p\,\JZ^q\,\KP^r}{Nj_0,z}, \label{eq:CGExpectation}
\end{eqnarray}
	where operators on the left hand side are defined via their fine descriptions \eqref{eq:DefCoarseKM} and they are simply given by the generators in representation $Nj$ on the right hand side. Furthermore, the probabilities to obtain a certain eigenvalue of $\JZ$ at the coarse and fine levels agree. 
	\item Since the Hamiltonian will be given by a linear combination of the generators, dynamics can be transferred between the coarse and fine descriptions due to \eqref{eq:groupOnCohStates}. In other words, computing time evolution as an action on the spinor $z$ commutes with coarse graining.

\end{itemize}

\subsection{Renormalised Hamiltonian from canonical quantisation}

While the coarse graining map from the previous section is computable analytically and can be extended to other groups \cite{BHAddendum}, it offers limited insights into quantum systems without such a group structure. In the example of loop quantum cosmology, one usually considers the elementary operators $\hat v$ and $\widehat{ e^{i \lambda b}}$ acting on a Hilbert space spanned by the eigenstates of $\ket{v}$ of $\hat v$ with eigenvalue $v$, where $\widehat{ e^{i \lambda b}}$ acts as a finite shift operator in the $v$-label. Then, one assembles operators corresponding to the Hamiltonian or $\kpm$ from these elementary operators. In contrast, the group quantisation procedure from section \ref{sec:GroupQuantisation} directly quantises $k_\pm$, so that access to its supposed constituents $\hat v$ and $\widehat{ e^{i \lambda b}}$ is lost in the process. As a consequence, the renormalisation group flow for the Hamiltonian operator in the group quantised model can be defined as ``change representation to $Nj_0$'', but the resulting operator could not be expressed via the elementary operators $\hat v$ and $\widehat{ e^{i \lambda b}}$ which would have offered an insight into how general Hamiltonians used in loop quantum gravity behave under coarse graining. 

\begin{figure}[h!]
	\centering
		\includegraphics[trim = 0mm 6cm 8cm 3cm, clip, scale=0.8]{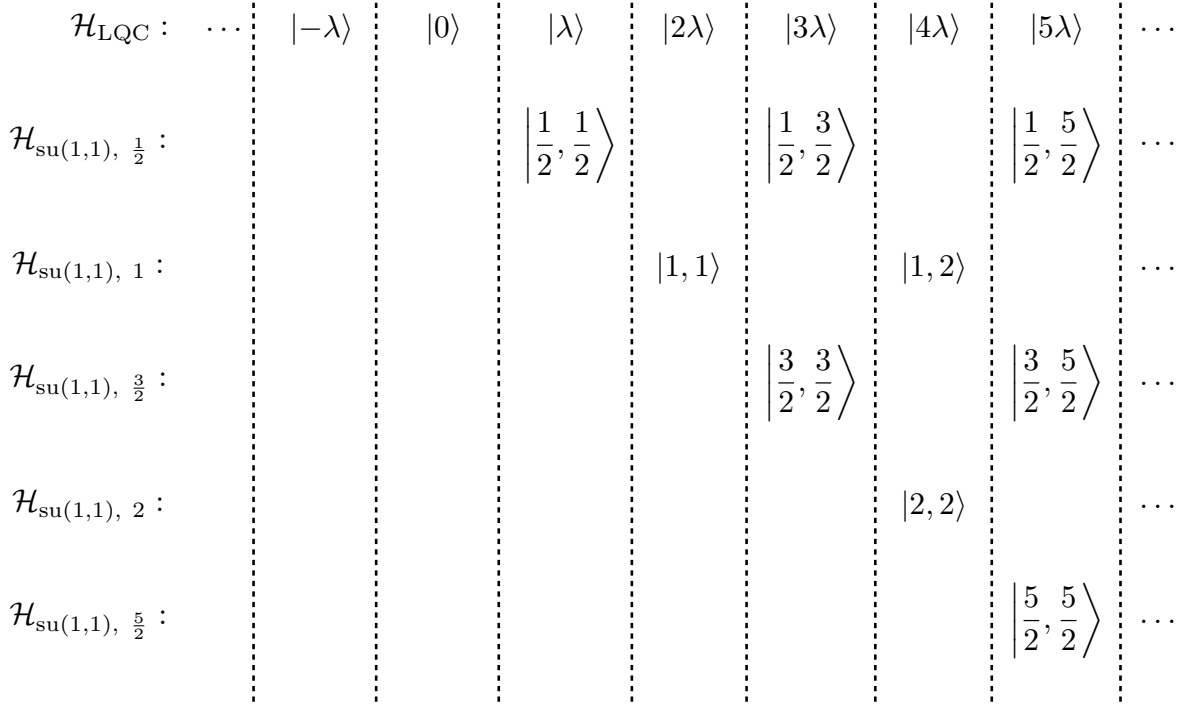}

	\caption{The embedding of the su$(1,1)$ representation spaces into the loop quantum cosmology Hilbert space derived in \cite{BWI} is shown. States $\ket{j,m}$ in an su$(1,1)$ representation space $\mathcal H_{\text{su}(1,1),~j}$ with label $j$ are mapped to the loop quantum cosmology state $\ket{2 \lambda m}$ shown in the top row of the same column. Due to the properties of the Hamiltonian operator (shift operators occur as squares), subspaces of $\mathcal H_\text{LQC}$ with volume eigenvalues containing only even or odd multiples of $\lambda$ are preserved by the dynamics. Moreover, since the Hamiltonian at coarseness level $j$ is constructed from su$(1,1)$ generators in representation $j$, it does not produce states with eigenvalues lower than $j$. Hence, the dynamics at coarseness level $j$ preserves the su$(1,1)$ representation space with label $j$, as required for consistency.  }\label{fig:Embedding}
\end{figure}

A strategy to overcome this was presented in  \cite{BWI}. The main idea was to assemble operators $\jz, \kpm$ from the elementary operators $\hat v$, $\widehat{ e^{i \lambda b}}$ directly on the loop quantum cosmology Hilbert space which satisfy the su$(1,1)$ Lie algebra relations for a given representation label $j$. This computation was carried out and operators for each $j \in \mathbb N/2$ were found. 
Thereby, one automatically obtained embeddings of the su$(1,1)$ representation spaces into the loop quantum cosmology Hilbert space and operators corresponding to the generators in arbitrary representations. Figure \ref{fig:Embedding} clarifies this embedding. 
As a consequence of \eqref{eq:GeneratorsH}, one can then read off the coarse grained Hamiltonian in representation $j$, which is interpreted as a renormalised Hamiltonian at coarseness scale $j=Nj_0$, i.e. with minimally resolved volume $v_m=2 \lambda j$. The expression is given by 
\begin{align} \label{eq:HRen}
	\hat H_g^{(j)} =& -\frac{1}{2 \lambda^2} \osin ~\sqrt{\hat v^2 - 4 \lambda^2 (j-1/2)^2} ~ \osin  \\
	&+  \frac{1}{8 \lambda^2} \BR{\sqrt{(\hat v +  \lambda)^2 - 4 \lambda^2 (j-1/2)^2}+\sqrt{ (\hat v -  \lambda)^2 - 4 \lambda^2 (j-1/2)^2}- 2 \hat v} \nonumber
\end{align}
and features corrections which are most relevant when the volume of the system is close to the minimal volume $2\lambda j$. As a consequence, it was shown in \cite{BWI} that the physics of many small spins (volumes) disagrees with that of few large spins that is usually captured by so-called effective equations \cite{AshtekarQuantumNatureOfAnalytical}. 

The main content of the present paper will be to compute the analogue of \eqref{eq:HRen} in a coherent state path integral, i.e. to extract the renormlised Hamiltonian, canonical action, and path integral measure directly from a coherent state path integral based on the group quantisation given in section \ref{sec:GroupQuantisation} or equivalently the canonical theory in \cite{BWI} sketched in this section.

\section{The SU(1,1) Coherent State Path Integral} \label{chap:cspi}

The path integral formalism due to Feynman is a well-known alternative to canonical quantisation that has found use in many areas of physics, but it has been particularly successful in high-energy physics. In typical derivations (e.g. \cite{SchulmanTechniquesAndApplications}), one introduces the path integral in the configuration space of some quantum mechanical system, which then straight-forwardly generalises to field theories. A different approach -- and less familiar in high-energy applications -- is to use coherent state path integrals (CSPI). The idea is to first introduce coherent states for the system under consideration, and then to apply all the usual steps in the derivation of the path integral formalism there. Such path integrals are more commonly used in condensed matter theory than in high-energy physics, but there are attempts to apply these techniques to Loop Quantum Gravity as well \cite{HanEffectiveDynamicsFrom}. 

In particular, looking at our work so far, we would like to derive a path integral formula using the Perelomov coherent states we have been using. If we succeed in this, we should find that the coarse graining property discussed in section \cref{sec:CG} is somehow realised in our result. We will not only see that this is indeed the case, but also that we can translate our CSPI result into phase space, where we can independently verify the result and better utilise our physical intuition, perhaps even to study the system numerically. First however, we need to set up a CSPI formula. To this end, we will first recall the basics of the formalism in \cref{sec:CSPIrecap}. Then, we will turn to the derivation of the CSPI for our model in \cref{sec:CSPIsu11}. A technical sublety which arises in that process will be dealt with in \cref{sec:identity}.

\subsection{Recapitulation of Coherent State Path Integrals}\label{sec:CSPIrecap}
We will begin by deriving the coherent state path integral for a particularly simple example: the harmonic oscillator. In this way, we can see the steps one usually follows to arrive at a CSPI formula, which we can then try to repeat for our cosmological system. Afterwards, we will also briefly note some potential problems of the coherent state path integral formalism once one treats slightly more complicated theories.

\subsubsection{The Harmonic Oscillator}\label{subsec:HO}
A useful review of the following well-known derivation can be found e.g. in \cite{BergeronCoherentStatePath}. We consider the quantum harmonic oscillator, with the ladder operators \( \oa,\oad \) satisfying
\begin{equation}
	\comm{\oa}{\oad}=1\punkt
\end{equation}
The associated Hilbert space is spanned by the states
\begin{equation}
	\ket{n}=\frac{1}{\sqrt{n!}}\left(\oad\right)^n\ket{0}\komma\quad\oa\ket{0}=0\komma
\end{equation}
where $\ket{0}$ is the ground state. Then, we define the usual coherent states for the harmonic oscillator, which are eigenstates of $\oa$:
\begin{equation}\label{eq:HOcohStates}
	\ket{z}=\e{z\oad}\ket{0}\komma\quad\oa\ket{z}=z\ket{z}\komma
\end{equation}
where $z\in\IC$. To find a coherent state path integral formalism, we will generally need
\begin{enumerate}
	\item the overlap between two coherent states,
		\begin{equation}
			\braket{z}{z'}=\e{\bar{z}z'}\komma
		\end{equation}
	\item a resolution of the identity,
		\begin{equation}\label{eq:HOidentity}
			\id=\int\frac{\dd{z}\dd{\bar{z}}}{2\pi i}\e{-z\bar{z}}\dyad{z}{z}\komma
		\end{equation}
	\item a decomposition of arbitrary states in terms of coherent states,
		\begin{equation}\label{eq:HOdecomposition}
			\ket{\psi}=\int\frac{\dd{z}\dd{\bar{z}}}{2\pi i}\e{-\abs{z}^2}\psi(\bar{z})\ket{z}\punkt
		\end{equation}
\end{enumerate}
The latter is of course always available when one has a resolution of the identity. With these ingredients, we now turn to the derivation of the path integral formula. Typically, we will be interested in matrix elements of the time evolution operator \( \e{-iT\oper{H}(\oad,\oa)} \), where \( \oper{H}(\oad,\oa) \) is the normal ordered Hamiltonian of the system under consideration. In principle, this can also be something different from the simple harmonic oscillator.

Now, if we want to compute the probability amplitude to evolve from an initial state $\ket{i}$ to a final state $\ket{f}$, the matrix element we are interested in is
\newcommand*{\myLim}{\lim_{\substack{M\to\infty\\T\text{~fixed}}}}
\begin{equation}
	\mel**{f}{\e{-iT\oper{H}(\oad,\oa)}}{i}\overset{\Delta t=\frac{T}{M+1}}{=}
	\myLim\mel**{f}{\left(1-i\Delta t\oper{H}(\oad,\oa)\right)^{M+1}}{i}\komma
\end{equation}
where we simply rewrote the exponential with a standard identity. In the limit $\Delta t\to0$, this expression can be understood as evolving the initial state $\ket{i}$ by successive, infinitesimal time steps rather than by a finite time $T$ all at once. We proceed as usual by inserting a resolution of the identity \eqref{eq:HOidentity} at each of these ``time slices'':
\begin{align}
	\mel**{f}{\left(1-i\Delta t\oper{H}(\oad,\oa)\right)^{M+1}}{i}&=
	\int\left(\prod_{j=1}^{M}\frac{\dd{z_j}\dd{\bar{z}_j}}{2\pi i}\right)\e{-\sum_{j=1}^{M}\abs{z_j}^2}\left[\prod_{k=1}^{M-1}\mel**{z_{k+1}}{\left(1-i\Delta t\oper{H}(\oad,\oa)\right)}{z_k}\right]\nonumber\\
	&\hphantom{=\int}\times\mel**{f}{\left(1-i\Delta t\oper{H}(\oad,\oa)\right)}{z_M}
	\mel**{z_1}{\left(1-i\Delta t\oper{H}(\oad,\oa)\right)}{i}
\end{align}
We can now get rid of the operators by using the property \eqref{eq:HOcohStates}: 
\begin{equation}\label{eq:HOhamiltonFunction}
	\mel**{z_{k+1}}{\left(1-i\Delta t\oper{H}(\oad,\oa)\right)}{z_k}=\braket{z_{k+1}}{z_k}\left[1-i\Delta t H(\bar{z}_{k+1},z_k)\right]
\end{equation}
Here, $H(\bar{z},z')$ is the function given by the matrix element \( \frac{\mel**{z}{\oper{H}}{z'}}{\braket{z}{z'}} \) for a normal-ordered Hamiltonian. Due to \cref{eq:HOcohStates}, it can be obtained by simply replacing all occurrences of $\oad$ with $\bar{z}$ and $\oa$ with $z$.

So far, everything we did was quite general and only contingent on the three requirements above being met. The trick used in eq. \eqref{eq:HOhamiltonFunction} however only works for the harmonic oscillator algebra, so we will need to find some alternative in our \( \al{su}(1,1) \) model. Plugging this into the transition amplitude, we find
\begin{equation}
	\begin{aligned}
		&\mel**{f}{\e{-iT\oper{H}(\oad,\oa)}}{i}=\\
		&\myLim\int\left(\prod_{j=1}^{M}\frac{\dd{z_j}\dd{\bar{z}_j}}{2\pi i}\right)
		\e{-\sum_{j=1}^{M}\abs{z_j}^2}\e{\sum_{j=1}^{M-1}\bar{z}_{j+1}z_j}
		\prod_{k=1}^{M-1}\left[1-i\Delta t H(\bar{z}_{k+1},z_k)\right]\\
		&\hphantom{\myLim\int}\times\braket{f}{z_M}\braket{z_1}{i}
		\left[1-i\Delta t\frac{\mel**{f}{\oper{H}}{z_M}}{\braket{f}{z_M}}\right]
		\left[1-i\Delta t\frac{\mel**{z_1}{\oper{H}}{i}}{\braket{z_1}{i}}\right]\punkt
	\end{aligned}
\end{equation}
To take the continuum limit, we define paths by the prescription $z_j=z(j\Delta t)$ and assume that they are \emph{continuous}:
\begin{equation}
	z_{j\pm1}=z(j\Delta t)\pm\dot{z}(j\Delta t)\Delta t
\end{equation}
In the \emph{symplectic} (or Berry phase) term \( z_j(\bar{z}_{j+1}-\bar{z}_j) \), this leads to the appearance of a time derivative, while the Hamiltonian symbol is approximated as ``diagonal'', i.e. by taking the expectation value at a single time slice, rather than the matrix element between adjacent slices:
\begin{equation}
	H(\bar{z}_{k+1},z_k)=H(\bar{z}_k,z_k)+\order{\Delta t}
\end{equation}

Finally, we expand the initial and final states in terms of coherent states according to \cref{eq:HOdecomposition} and take the continuum limit $\Delta t\to0$, yielding our final result:
\begin{equation}\label{eq:HOpathIntegral}
	\begin{aligned}
		\mel**{f}{\e{-iT\oper{H}(\oad,\oa)}}{i}&=\int\DD{z}\DD{\bar{z}}
		\e{i\int_{0}^{T}\dd{t}\left[\frac{z\partial_t\bar{z}-\bar{z}\partial_t z}{2i}-H(\bar{z},z)\right]}\\
		&\hphantom{=\int\DD{z}}\times
		\e{\einhalb(\abs{z_i}^2+\abs{z_f}^2)}\bar{\psi}_f(z_f)\psi_i(\bar{z}_i)\\
		&\eqqcolon\int\DD{\myZ}~\,\e{\einhalb(\abs{z_i}^2+\abs{z_f}^2)}\bar{\psi}_f(z_f)\psi_i(\bar{z}_i)
	\end{aligned}
\end{equation}
Here, \( z_{i,f},\psi_{i,f} \) are the initial/final coordinates and wave functions respectively. The important takeaway is that this result splits up into some projection terms that fix the boundary conditions on the one hand, and a quantity $\DD{\myZ}$, which we will call the \emph{path integral kernel}. The task of \cref{sec:CSPIsu11} will be to identify this kernel for the \( \al{su}(1,1) \) Perelomov coherent states. 

\subsubsection{Potential Problems: The Bose-Hubbard Model}\label{subsec:problems}
The derivation in the previous section is quite simple, and at first sight, it seems to work for arbitrary systems described by a Hamiltonian that can be written in terms of the Heisenberg algebra \( \mathfrak{h}=\text{span}\{\id,\oa,\oad,\oad\oa=\oper{n}\} \). Unfortunately, it turns out not to be quite so easy, cf. e.g. \cite{WilsonBreakdownOfThe}. To see this, let us consider the one-site Bose-Hubbard model, given by the normal-ordered Hamiltonian
\begin{equation}
	\oper{H}=-\mu\oper{n}+\frac{U}{2}\oper{n}(\oper{n}-1)=-\mu\oad\oa+\frac{U}{2}\oad\oad\oa\oa\punkt
\end{equation}
Now, we can perform a Wick rotation $t\to i\tau$ in the above path-integral formula to obtain the path integral version of the statistical partition function \cite{AtlandBook}, and then simply plug the Bose-Hubbard Hamiltonian, normal ordered as per the requirements above, into \eqref{eq:HOpathIntegral}, simply replacing $\oad$ with $\bar{z}$ and $\oa$ with $z$ as before. This yields
\begin{equation}
	\mathcal{Z}'=\int\DD{z}\DD{\bar{z}}\e{-\int_{0}^{\beta}\dd{t}\left[\frac{z\partial_t\bar{z}-\bar{z}\partial_t z}{2i}-\mu\abs{z}^2+\frac{U}{2}\abs{z}^4\right]}\punkt
\end{equation}
In this case, we can even solve this path integral exactly, using the method described in \cite{WilsonBreakdownOfThe}. The partition function is then found to be
\begin{equation}
	\mathcal{Z}'=\sum_{n=0}^{\infty}\e{\mu n\beta-\frac{U}{2}n^2\beta}\komma
\end{equation}
while the (correct) statistical mechanical calculation of the partition function gives
\begin{equation}
	\mathcal{Z}=\tr\e{-\beta\oper{H}}=\sum_{n=0}^{\infty}\e{\mu n\beta-\frac{U}{2}n(n-1)\beta}\komma
\end{equation}
so the two results do not agree. The natural question then is: where and when does this method go wrong? Clearly, for highly excited systems ($n\gg1$), the two methods will agree, so the coherent state path integral is at least an approximation to the exact result, and it seems it might be possible to fix our approach. 

Firstly, we should mention that this problem does not appear in the cosmological system studied in this work. It turns out only to make a difference in \emph{interacting} theories, where the Hamiltonian contains terms non-linear in the generators used to construct the coherent states. In our case, this generator is $\kp$, which appears only linearly in the Hamiltonian. So for the moment, we can simply ignore this complication. However, let us note that interacting theories are necessary as soon as one considers scalar fields with (inflationary) potentials, see the discussion in \cite{HanederMasterThesis}. 

Secondly, the incongruence of the results -- which stems from unwarranted continuity assumptions -- can be avoided if we strictly work with the discrete version of the path integral and do not take the continuum limit that is performed in \cref{subsec:HO}.
However, analytical calculations often rely on the continuous form, and were we to include inflationary potentials for the field, we might very well have to include non-linear terms in the Hamiltonian. These two factors put together make it clear that a genuine solution to this problem is in order. Luckily, such a solution exists \cite{BruckmannRigorousConstructionOf}.

To understand how it works, we first need to pinpoint where the problems are coming from; namely, we used two unjustified approximations in the above derivation: first, that the Hamiltonian symbol appearing in \cref{eq:HOhamiltonFunction} is approximately diagonal,
\begin{equation}
	H(\bar{z}_{k+1},z_k)=H(\bar{z}_k,z_k)+\order{\Delta t},
\end{equation}
i.e. the assumption of \emph{continuous paths}; and second\footnote{This is only the case in the solution of the Bose-Hubbard model path integral, cf. \cite{WilsonBreakdownOfThe}.}, an erroneous assumption that the discrete version of the symplectic term can be transformed to polar coordinates as
\begin{equation}
	\bar{z}_k(z_{k+1}-z_k)=in_k(\theta_k-\theta_{k-1})+\order{\Delta t}\punkt
\end{equation}
These errors are corrected in two ways: first, one works with the Glauber-Sudarshan P-representation \cite{ScullyBook} of the Hamiltonian,
\begin{equation}
	\oper{H}=\int\frac{\dd[2]{z}}{2\pi i}h(\bar{z},z)\dyad{z}{z}\komma
\end{equation}
and uses the explicitly diagonal $h$-symbol instead of $H(z,z')$. Second, one transforms the off-diagonal part of the symplectic term in a way in which terms on adjacent time slices can be factorised. In this way, the variables on different time slices can be integrated independently, and no continuity assumptions are necessary.

We see then that coherent state path integrals can be used even in non-linear systems, provided one takes care to define them correctly. Knowing this, we may embark on the application of this method to our cosmological system, assured that even the inclusion of realistic matter fields and non-linear potentials can be handled and will not break the formalism.

\subsection{Derivation of the SU(1,1) Coherent State Path Integral for Loop Quantum Cosmology}\label{sec:CSPIsu11}

\subsubsection{Coherent state path integral}

Let us begin then in the same way we did for the simple example of the Heisenberg algebra: we collect the required ingredients, then turn to the derivation of the path integral kernel. We already know the coherent states:
\begin{equation}
	\ket{j,z}=(2L(z))^j\sum_{m=j}^{\infty}\sqrt{\frac{\Gamma(m+j)}{\Gamma(m-j+1)\Gamma(2j)}}
	\frac{(z_1)^{m-j}}{(\bar z_0)^{m+j}}\ket{j,m}\tag{\ref{eq:cohStates}}
\end{equation}
Looking back at our requirements in \cref{subsec:HO}, we need the overlap between two coherent states. This can be straightforwardly calculated as
\begin{equation}\label{eq:cohStatesOverlap}
	\braket{j,x}{j',y}=\delta_{j,j'}\frac{(2L(x))^j(2L(y))^j}{(x_0\bar{y}_0-\bar{x}_1y_1)^{2j}}\punkt
\end{equation}
We also already mentioned a resolution of the identity,
\begin{equation}
	\id=\int\dd[2]{z_0}\dd[2]{z_1}\frac{(2j-1)}{2\ell\pi^2}\delta(2L(z)-2\ell)\dyad{j,z}\komma\tag{\ref{eq:identityResolution}}
\end{equation}
along with a precise definition of it in the case $j=\nicefrac{1}{2}$ in appendix \ref{sec:identity}.

For now though, let us turn to the derivation of the path integral kernel. In principle, we still need a decomposition of arbitrary states in terms of coherent states. However, it should be obvious that with the identity operator \eqref{eq:identityResolution}, such a decomposition can be trivially found as \( \ket{\psi}=\id\ket{\psi} \). 

We begin then in the same way as before: we want to compute the transition amplitude from one state to another, and ``slice up'' the time evolution operator in $M$ time slices separated by a time increment $\Delta t=\frac{T}{M+1}$. However, our way of slicing the operator is slightly different this time. Suppressing the label $j$, we have
\begin{equation}\label{eq:TransitionAmplitude}
	\mel**{f}{\e{-i\oper{H}T}}{i}=\myLim\mel**{f}{\left(\e{-i\oper{H}\Delta t}\right)^{M+1}}{i}\punkt
\end{equation}
Note that this step is also correct if we leave $M$ finite. As before, we insert a resolution of the identity at every time slice to obtain
\begin{equation}\label{eq:insertIdentity}
	\begin{aligned}
		\mel**{f}{\left(\e{-i\oper{H}\Delta t}\right)^{M+1} }{i}&=
		\int\left(\prod_{m=1}^{M}\dd[4]{z_m}\frac{2j-1}{2\ell\pi^2}\delta(2L(z_m)-2\ell)\right)
		\left[\prod_{k=1}^{M-1}\mel**{z_{k+1}}{\e{-i\oper{H}\Delta t}}{z_k}\right]\\
		&\hphantom{=\int}\times\mel**{f}{\e{-i\oper{H}\Delta t}}{z_M}\mel**{z_1}{\e{-i\oper{H}\Delta t}}{i}\punkt
	\end{aligned}
\end{equation}
where \( \dd[4]{z}=\dd[2]{z_0}\dd[2]{z_1} \) and $z_k$ refers to the \emph{spinor} on the $k$-th time slice. In this form, it should be clear why we elected to slice up the time evolution operator in this way. Namely, we know exactly how these sliced time evolution operators -- which are after all elements of \( \gr{SU}(1,1) \) -- act on the coherent states: as a simple multiplication on the spinor. Hence, we can write formally:
\begin{equation}\label{eq:TimeEvolvedOverlap}
	\mel**{z_{k+1}}{\e{-i\oper{H}\Delta t}}{z_k}=\braket{z_{k+1}}{\e{-iH\Delta t}z_k}\eqqcolon\braket{z_{k+1}}{z_{k+\Delta t}}
\end{equation}
We can plug this explicit time evolution back into \cref{eq:insertIdentity} and bring our amplitude into the form
\begin{equation}
	\mel**{f}{\e{-i\oper{H}T}}{i}=\int\DD[4]{z}\e{iS}\times\text{projection~terms}=\int\DD{\myZ}\times\text{projection~terms}\komma
\end{equation}
where we once again identify the kernel $\DD{\myZ}$.

Let us first discuss the action. To find the corresponding terms, we simply use the overlap \eqref{eq:cohStatesOverlap}, as well as the known time evolution \eqref{eq:zOfT}. Renaming $\Delta t\to2\lambda\epsilon$, i.e. introducing the dimensionless time increment $\epsilon$, we get from each factor of \eqref{eq:TimeEvolvedOverlap} in \eqref{eq:insertIdentity} a contribution of 
\begin{equation}
	\begin{aligned}
	i L_{k+1,k} := \log\braket{z_{k+1}}{\e{-iH\Delta t}z_k}&=j\Biggl(-2 \log\left( z_{0,k+1} \bar{z}_{0,k} - z_{1,k} \bar{z}_{1,k+1} \right) + 2 \log \left( 2 \ell \right) \\
	&\hphantom{=j\left\{\right\}}+i\epsilon\frac{\left(\bar{z}_{0,k}-z_{1,k}\right)\left[2\ell+\bar{z}_{1,k+1}(z_{1,k+1}-\bar{z}_{0,k+1})\right]}{-z_{1,k}\bar{z}_{1,k+1}\bar{z}_{0,k+1}+\bar{z}_{0,k}(2\ell+\abs{z_{1,k+1}}^2)}
	\vphantom{\left(\frac{\bar{z}_{0,k}(2\ell+\abs{z_{1,k+1}}^2)}{\bar{z}_{0,k+1}}\right)}\Biggr)+\order{\epsilon^2}
	\end{aligned}
\end{equation}
to $iS$. We note the very simple dependence on the representation label/extensive scale $j$ -- a direct proportionality. This is a consequence of the fact that $j$ only appears in exponents in the overlap of two coherent states, and that the action of the time evolution operator is \emph{always} completely absorbed into the spinor, irrespective of the irrep. Hence, the action, which is simply a sum of logarithms of such overlaps, will be exactly proportional to $j$.

To carry out the continuum limit, we further simplify our expressions by assuming that we have \emph{continuous paths}, i.e.
\begin{equation}\label{eq:continuumLimit}
	z_k\to z(t_k)\komma\quad z_{k\pm1}\to z(t_k)\pm\dot{z}(t_k)\Delta t
\end{equation}
in the limit $\Delta t\to0$. As we recall from \cref{subsec:problems}, this assumption is not justified in general and may lead to wrong results. However, due to the simple Hamiltonian, linear in the generators, we do not expect any problems in our case. As a crosscheck, it was verified that the equations of motion computed from the continuum spinor action are solved by the known time evolution \eqref{eq:zOfT} \cite{HanederMasterThesis}. Hence, it is enough at this stage to take the continuum limit at the level of the action. Inserting \eqref{eq:continuumLimit} into $i L_{k+1,k}$  we get
\be
	i L_{k+1,k} = - i j  \epsilon \frac{\bar{z}_0^2 \left( -4 i \lambda \dot{z}_0  + \bar{z}_1 \right) + z_1 \left( 2 \ell +|z_1|^2 \right) -2 \bar{z}_0 \left( \ell + |z_1|^2 -2 i \lambda z_1 \dot{\bar{z}}_1  \right) }{2 \ell \bar{z}_0}(t_k) +\order{\epsilon^2} \text{,}
\ee
noting that the constant term vanishes. In the continuum limit, the sum over such terms becomes an integral as $ \sum\epsilon\xrightarrow{\epsilon\to0}\int\frac{\dd{t}}{2\lambda} $, leading to 
\be
	i S = - i j \int_{0}^{T}\dd{t}  \frac{\bar{z}_0^2 \left( -4 i \lambda \dot{z}_0  + \bar{z}_1 \right) + z_1 \left( 2 \ell +|z_1|^2 \right) -2 \bar{z}_0 \left( \ell + |z_1|^2 -2 i \lambda z_1 \dot{\bar{z}}_1  \right) }{4 \lambda \ell \bar{z}_0}  \text{.} \label{eq:ActionContinuumSpinor}
\ee

We now turn to the integration measure and introduce polar coordinates for each time step as $z_0 = |z_0| e^{i\phi_0}$ and $z_1 = |z_1| e^{i\phi_1}$, leading to 
\begin{align}
	& \int_{|z_0|^2 \geq 2 \ell}\dd[2]{z_0} \int_{|z_1^2|\leq |z_0|^2}\dd[2]{z_1} \frac{2j-1}{2\ell\pi^2} \delta(2L(z)-2\ell) \\
	=& \int_{2 \ell}^\infty \dd{|z_0|^2}\int_{0}^{|z_0|^2} \dd{|z_1|^2} \int_{0}^{2\pi}\dd{\phi_0}\int_{0}^{2\pi}\dd{\phi_1} \frac{2j-1}{8\ell\pi^2} \delta(|z_0|^2-|z_1|^2-2\ell)
\end{align}
The integrals over $|z_1|^2$ in \eqref{eq:insertIdentity} can be explicitly performed by dropping the $\delta$-distribution and substituting $|z_1|^2 \rightarrow |z_0|^2-2 \ell$.

\subsubsection{Coarse graining in the path integral}

In order to talk about coarse graining in the path integral setting, we need to define a path integral for the fine grained system as well. This is very simple: once again, we compute the transition amplitude from an initial state $\ket{i}$ to a final state $\ket{f}$, analogously to \cref{eq:TransitionAmplitude}, slice up the time evolution operator, and insert an identity at every time slice. Now however, instead of projecting with some $\dyad{Nj_0,z}$, we use a product state $\ket{j_0,z_1}\otimes\ket{j_0,z_2}\otimes\ldots$, i.e. the same state as in the Hamiltonian formalism, \cref{eq:ProductState}\footnote{The identity is then simply defined as a tensor product of $N$ identities in representation $j_0$.}. Likewise, we express the Hamiltonian using \cref{eq:DefCoarseKM} and its analogues for the other generators to get an operator acting on the product states. From here on, it is possible to simply repeat all of the above steps, and one quickly notices that the path integral completely factorises into contributions of each subsystem. 

Let us look in more detail at how the coarse graining property introduced in \cref{sec:CG} is realised in this path integral formalism. We recall that physically, coarse graining means that we consider a cosmological system described by $N$ subsystems, all with a minimal volume $2\lambda j_0$, where $j_0$ is the label of the irreducible \( \al{su}(1,1) \) representation acting on the Hilbert space of a subsystem. What we want to do then is to faithfully describe the coarse properties of this setup in terms of a single coarse grained system with minimal volume $2\lambda Nj_0$. To do this, we need to determine how the path integral -- meaning the action and the measure -- renormalises under coarse graining. 

In particular, we expect to find the coarse graining property realised in matrix elements of \( \al{su}(1,1) \) operators, of the form
\begin{equation}\label{eq:CSPImatrixElements}
	\ev{\oper{\mathcal{O}}}=
	\frac{\displaystyle\mathcal{N}_j\int\DD[2]{z_0}\DD[2]{z_1}\e{iS[z]}\mathcal{O}}{\displaystyle\mathcal{N}_j\int\DD[2]{z_0}\DD[2]{z_1}\e{iS[z]}}\punkt
\end{equation}
Here, we can immediately see that the $j$-dependent normalisation factor $\mathcal{N}_j$ cancels out between numerator and denominator in expectation values (and indeed in all matrix elements, where the only difference would be the inclusion of suitable projection terms in the path integral). Therefore, we do not need to examine how it renormalises under coarse graining, and only the measure and the action are of interest. Looking at the latter, we already noted that the right-hand side of \cref{eq:ActionContinuumSpinor} is exactly proportional to $j$. Hence, the action is an extensive quantity, and we find that comparing the coarse grained system to the fine grained system, the actions are respectively described by
\begin{equation}\label{eq:actionRenormalisation}
	S_{Nj_0}[z]=N\cdot S_{j_0}[z]\text{~vs.~} S_{j_0\otimes j_0\otimes\ldots}[z_1,z_2,\ldots,z_N]=\sum_{i=1}^{N}S_{j_0}[z_i]
\end{equation}
This means that the only function we need to path-integrate is the simple action $S_{j_0}$ of a single system at some coarse graining scale $j_0$ in both cases. There are no more complicated structures arising from the coarse graining transformation. However, these two actions are not simply identical, because the spinors in the fine grained system are still independent. Hence, we can not simply write \( S_{Nj_0}=S_{j_0\otimes j_0\otimes\ldots} \). This property is analogous to the spin foam model analyzed in \cite{HanSemiclassicalBehaviorOf}.

Now let us look at the measure. As already mentioned, we do not have to care about the normalisation factor $\mathcal{N}_j$. Meanwhile, the rest of the measure does not depend on $j$ at all. Thus, the renormalisation is extremely simple; we merely get one copy of the measure for each subsystem:
\begin{equation}\label{eq:measureRenormalisation}
	\DD[2]{z_0}\DD[2]{z_1}\text{~vs.~}\prod_{i=1}^{N}\DD[2]{z_{0,i}}\DD[2]{z_{1,i}}
\end{equation}

Putting these two results together, we conclude that the path integral of a collection of $N$ subsystems at coarse graining scale $j_0$ \emph{factorises exactly} into $N$ path integrals, each describing a single system. For the measure, this is simply the content of \cref{eq:measureRenormalisation}. Meanwhile, using \cref{eq:actionRenormalisation}, we find that \( \e{iS_{j_0\otimes j_0\otimes\ldots}}=\e{iS_{j_0}[z_1]}\e{iS_{j_0}[z_2]}\ldots\e{iS_{j_0}[z_N]} \). Hence, at the level of the path integral kernel, we find
\begin{equation}
	\DD{\myZ_{Nj_0}[z]}\text{~vs.~}\prod_{i=1}^{N}\DD{\myZ_{j_0}[z_i]}
\end{equation}
Of course, physically, this is hardly surprising, and like in the Hamiltonian formalism, simply an expression of the fact that the systems are completely uncorrelated, and hence the path integral \emph{should} factorise into single-system contributions. However, as has been discussed in  \cite{BHAddendum}, there is an underlying group theoretical reason why this works. It may prove an interesting avenue for further study to transfer this argument to the path integral setting, and possibly find some more general criteria for when coarse graining works exactly at this level as well.

\subsubsection{Phase space path integral}

Next, we would like to translate the path integral to phase space. For this, we need a map from the remaining spinor variables to $v,b$ as well as a further variable $\varphi$ that will turn out to be irrelevant for the phase space description, i.e. encode a symmetry transformation. Such a transformation is e.g. given in \cite{BenAchourThiemannComplexifierIn}, but we will start with a more general ansatz here to motivate the result. We have two requirements: 
\begin{enumerate}
	\item We would like to obtain an integration measure of the from $\int \dd v \int \dd b \int \dd \varphi$, i.e. with trivial kernel up to a constant. For this, the Jacobian matrix of the transformation $(|z_0|^2, \phi_0, \phi_1) \rightarrow (v, b, \varphi)$ should have constant non-zero determinant. Moreover, the range of $\lambda b$ should be restricted to $[0,2\pi]$.
	\item When inserting observables into the phase space path integral, we would like that $v$ and $b$ correspond to their classical counterparts. This requires that the expectation values of of $\jz, \kpm$ in coherent states reduce to their classical counterparts \eqref{eq:ClassicalGenerators} when the spinors are written as functions of $v,b,\varphi$. 
\end{enumerate}
We start by observing from \eqref{eq:Ejz} that $v$ is encoded in the modulus of the spinors, whereas a comparison of \eqref{eq:Ekp} and \eqref{eq:ClassicalGenerators} shows that $b$ is encoded in their phases. This suggests the ansatz
\begin{equation}
	\begin{aligned} 
		z_0(t)&=f_0(v)~ \e{-i\lambda b(t)+i\varphi(t)}\komma&&
		\bar{z}_0(t)=f_0(v)~\e{i\lambda b(t)-i\varphi(t)}\komma\\
		z_1(t)&=f_1(v)~\e{-i\lambda b(t)-i\varphi(t)}\komma&&
		\bar{z}_1(t)=f_1(v)~\e{i\lambda b(t)+i\varphi(t)},
	\end{aligned}
\end{equation}
where $f_0$ and $f_1$ are two non-negative functions. The constraint $|z_1|^2 = |z_0|^2-2 \ell$ immediately yields $f_1^2 = f_0^2-2\ell$. Inserting into \eqref{eq:Ejz} and equating to $v/2\lambda$ now implies 
\be
	f_0^2(v) = \frac{\ell}{2 \lambda j} v + \ell \text{.}
\ee
Due to the linear dependence of $f_0^2$ on $v$, it follows that 
\begin{equation}
	J=\abs{\det(\pdv{(|{z}_0|^2,\phi_0,\phi_1)}{(v,b,\varphi)})}=\frac{\ell}{j},
\end{equation}
i.e. requirement $1$ is satisfied. Finally, we insert the ansatz into \eqref{eq:Ekp}, which implies
\be
	\sigma = (2 \lambda j)^2 = v_m^2 \text{.}
\ee	
To conclude, requirement 2 selects a unique choice for $\sigma$ as a function of $j$ and automatically satisfies requirement 1. The transformation to phase space variables thus reads
\begin{equation}\label{eq:spinorsToPhasespace}
	\begin{aligned} 
		z_0(t)&= \sqrt{\ell} \sqrt{\frac{v}{v_m}  + 1 } ~ \e{-i\lambda b(t)+i\varphi(t)}\komma&&
		\bar{z}_0(t)=\sqrt{\ell} \sqrt{\frac{v}{v_m}  + 1 }~\e{i\lambda b(t)-i\varphi(t)}\komma\\
		z_1(t)&=\sqrt{\ell} \sqrt{\frac{v}{v_m}  - 1 }~\e{-i\lambda b(t)-i\varphi(t)}\komma&&
		\bar{z}_1(t)=\sqrt{\ell} \sqrt{\frac{v}{v_m}  -1 }~\e{i\lambda b(t)+i\varphi(t)},
	\end{aligned}
\end{equation}
The integration ranges of the spinors imply that $v$ has to be integrated from $v_m$ to $\infty$, while $\lambda b$ and $\varphi$ are restricted to $[0,2\pi]$. This leads, again dropping constant terms, to the integration measure 
\begin{align} \label{eq:MeasureTimeStep}
	 \int_{v_m}^\infty \dd{v} \int_{0}^{2 \pi / \lambda} \dd{b} \int_{0}^{2\pi} \dd{\varphi}
\end{align}
for each time step. We note that $\ell>0$ is still completely arbitrary and will also in the rest of the paper drop from all physical results. 

We are now in a position to insert \eqref{eq:spinorsToPhasespace} into \eqref{eq:ActionContinuumSpinor} to obtain the continuum action in terms of phase space variables. We find 
\be \label{eq:ActionContinuumPhaseSpace}
	i S = i  \int_{0}^{T}\dd{t} \left( v \dot b -H_\text{PI}(v,b) -  \frac{v_m}{\lambda} \dot \varphi \right)
\ee
with 
\be
	H_\text{PI}(v,b) = -\frac{1}{2}\frac{\sin[2](\lambda b)}{\lambda^2}\sqrt{v^2-v_m^2}	+\frac{1}{4\lambda^2}\left(\sqrt{v^2-v_m^2}-v\right), ~~~ v_m = 2 \lambda j \label{eq:HPIvb}
\ee
We note as a consistency check that $\ell$ disappeared completely from the phase space expression of the path integral. The integral of the last term involving $\varphi$ in \eqref{eq:ActionContinuumPhaseSpace} can be performed explicitly and leads to a boundary term in the time integral. Thus, the integration over $\varphi$ in intermediate time steps can be dropped, i.e. absorbed as a constant term in the measure. At the initial and final time step, one can fix $\varphi$ via the boundary conditions and absorb this contribution into the measure. This leads to the final phase space path integral  
	\be \label{eq:phasespacePI}
		\int\DD{\myZ}=\mathcal{N}_j  \int \DD{v}  \DD{b}  \e{\displaystyle i S[v,b] }~~, ~~~~~~ S[v,b] = \int_{0}^{T}\dd{t} \left( \dot{b}(t){v}(t)-H_\text{PI}(v(t),b(t)) \right)
	\ee
with the integration boundaries at each time step as in \eqref{eq:MeasureTimeStep}. $\mathcal{N}_j$ stands pictorially for all constant terms that we have dropped and disappears in any matrix elements
\begin{equation}\label{eq:CSPImatrixElements}
	\ev{\oper{\mathcal{O}}}=
	\frac{\displaystyle\mathcal{N}_j \int \DD{v} \DD{b} \e{iS[v,b]}\mathcal{O}}{\displaystyle\mathcal{N}_j \int \DD{v} \DD{b} \e{iS[v,b]}}\komma
\end{equation}
where $\mathcal O$ can be any function of $v$ and $b$. Hence, we can also drop $\mathcal N_j$ from \eqref{eq:phasespacePI}. After this, the path integral contains no more pathology in the case $j=\nicefrac{1}{2} \Leftrightarrow v_m = \lambda$ and we can use it for all $j \in \mathbb N/2$.

\section{Discussion} \label{sec:Discussion}

Equation \eqref{eq:phasespacePI} provides us with the desired phase space path integral formulation of the group quantised cosmological model that was investigated before via group quantisation \cite{BodendorferCoarseGrainingAs, BHAddendum} and canonical quantisation \cite{BWI}. Due to the equivalence of the three existing formulations, all results concerning coarse graining directly transfer also to the phase space path integral derived in this paper. Hence, the conclusions drawn in \cite{BWI} apply as well. We will shortly reiterate them here and refer to \cite{BWI} for more details.
\begin{itemize}
	\item A realistic quantum state in full loop quantum gravity is usually thought of as a state involving many small quantum numbers (spins). In our case, this translates to states with small volume eigenvalues. Due to the formulation of our model with a fixed amount of fiducial cells, corresponding to graph-preserving regularisations in full loop quantum gravity, the quantum numbers cannot always be small throughout the evolution. It seems most plausible then to consider a realistic quantum state to have small quantum numbers close to the bounce that substitutes the big bang and big crunch singularities in the present model, where the volume is minimal.
	\item We assume that \eqref{eq:phasespacePI} defines the correct (most fine grained) dynamics of our model for the case $j_0=\nicefrac{1}{2}$. Plugging in the above ``many small quantum number'' state into the time evolution then yields the correct evolution of the universe. Due to the coarse graining property from section \ref{sec:CG}, it is also possible to describe the evolution via a quantum state involving a few large quantum numbers. Then however, one is forced to use the renormalised Hamiltonian with a correspondingly larger $j$ and larger $v_m = 2 \lambda j$. 
	\item It is common practice in loop quantum cosmology and loop quantum gravity to neglect renormalisation from small to large quantum numbers and simply use the analogue of the $j=\nicefrac{1}{2}$ Hamiltonian for states with large quantum numbers close to the bounce. In this case, the quantum evolution is well approximated by so-called effective equations for which the quantum Hamiltonians are used like classical expressions. For the model at hand, this leads to an overestimation of the critical density by a factor of $2$ as compared to the renormalised dynamics. The error made this way is very sensitive to taking the quantum numbers even slightly too large, i.e. if we peak the bounce volume on $2 v_m$ instead of $v_m$, we already overestimate the critical density by $87\%$ as opposed to $100\%$ at infinite bounce volume\footnote{To obtain these numbers, we simply evaluate the functions plotted in figure 1 of \cite{BWI} at $\alpha \approx 0.268$.}. 
	\item While the model system considered in this paper as well as \cite{BWI} is very simple and not claimed to be an accurate reflection of our universe, it serves the purpose of illustrating the importance of renormalisation. In more complicated systems, it would be very surprising if the effects of neglecting renormalisation would be less pronounced. In this sense, the conclusions of \cite{BWI} are not surprising, but merely illustrate expected facts in an analytically tractable toy model. 
\end{itemize}

While the conclusions of \cite{BWI} were mainly relevant for canonical formulations of loop quantum gravity, the present paper shows that they are also relevant for path integrals such as the spin foam models (see \cite{PerezTheSpinFoam} for an overview) developed as covariant formulations of loop quantum gravity. Investigations of renormalisation there (see \cite{SteinhausCoarseGrainingSpin} for an overview) were so far limited to the regime of large quantum numbers (spins), whereas the present paper along with the results of \cite{BWI} suggests that the effects are strongest for the coarse graining of the smallest to medium large quantum numbers. 

An interesting observation which contrasts the canonical \cite{BWI} and path integral quantisations \eqref{eq:phasespacePI} slightly is that the canonical Hamiltonian operator \eqref{eq:HRen} does not have any small volume corrections for $j=\nicefrac{1}{2}$. This implies that even if such corrections are absent in the fundamental (fine grained) Hamiltonian, they may occur as a result of renormalisation and even be large, i.e. $v_m$ is at the order of the smallest resolved volume in that case. However, it should also be noted that the ``absence'' of small volume corrections is a matter of factor ordering, and \eqref{eq:HRen} can also be reordered to an expression that involves small volume corrections even for $j=\nicefrac{1}{2}$. In this sense, ``absence of small volume corrections'' is not a well-defined property of a Hamiltonian operator of this kind.

This work opens a few interesting perspectives deserving to be further explored: one interesting direction is the relation to the full theory of Loop Quantum Gravity, either in the canonical formulation or in the spin foam formulation. It would be remarkable if the coarse graining operation discussed here can relate to the lattice refinement and renormalization in spin foam models \cite{BahrHypercuboidalRenormalizationIn,BahrNumericalEvidenceFor} or in the coherent path integral of canonical Loop Quantum Gravity \cite{HanEffectiveDynamicsFrom}. A question we ask is whether a fundamental description with small spins can lead to a better understanding of the refinement limit of the theory, given that small spins is a fine-grained description. On the other hand, it is important to go beyond the symmetry-reduced model and take into account nonhomogeneous and nonisotropic degrees of freedom. Hence, developing a cosmological perturbation theory is important to extract physical predictions from the formalism.

\section*{Acknowledgments}

NB was supported by an International Junior Research Group grant of the Elite Network of Bavaria. MH received support from the National Science Foundation through grant PHY-1912278.

\newpage 
\begin{appendix}

\section{Representation Theory of SU(1,1)}\label{app:SU11}
In this appendix, we review some useful facts about the group SU(1,1), the generalised special unitary group, and its Lie algebra, \( \al{su}(1,1) \) \cite{RamondBook}. The latter contains the three generators \( j_z,k_\pm \), which satisfy the algebra
\begin{equation}\label{eq:su11commutators}
	\comm{j_z}{k_\pm}=\pm k_\pm\komma\qquad \comm{k_+}{k_-}=-2j_z\komma
\end{equation}
which at first sight looks very similar to the algebra of the SU(2) generators, up to the sign in the second commutator. The similarity is also apparent when looking at the Casimir element of the algebra,
\begin{equation}\label{eq:su11casimir}
	\cas=j_z^2-\einhalb k_+k_--\einhalb k_-k_+=j_z^2-k_x^2-k_y^2=j(j-1)\komma
\end{equation}
where \( k_x=\frac{k_++k_-}{2},\,k_y=\frac{k_+-k_-}{2i} \). $j$ labels the representation of \( \al{su}(1,1) \) in which the Casimir is evaluated. Furthermore, we can write down a basis for \( \al{su}(1,1) \) in its defining representation,
\begin{equation}\label{eq:definingRep}
	j_z=\einhalb\pmqty{\pmat{3}}\komma\qquad k_x=\frac{i}{2}\pmqty{\pmat{2}}\komma\qquad k_y=-\frac{i}{2}\pmqty{\pmat{1}}\komma
\end{equation}
i.e. the basis are the Pauli matrices with suitable complex coefficients. The Pauli matrices themselves of course provide a basis for \(\al{su}(2) \).

Now given all these similarities between \( \al{su}(2) \) and \( \al{su}(1,1) \), one might be tempted to try to obtain irreducible representations (irreps) from the irreps of $\al{su}(2)$ in a similar way as one obtains the generators of the one algebra from those of the other. In this way, one would indeed obtain irreducible representations, but they would all fail to be unitary. In fact, as SU(1,1) is a non-compact group, \emph{all} unitary irreps are necessarily infinite dimensional. We can classify them by looking at the value of the Casimir, or equivalently the value of $j$, as well as the eigenvalues \( m \) of \( j_z \).

The first and (for us) most important series of irreps is the \emph{discrete series}. They are classified by a positive integer or half integer, \( j\in\nicefrac{\IN}{2} \), much like SU(2) irreps. The spectrum of $j_z$ is also discrete (in fact, this is always the case), and for each $j$, there are 2 distinct representations, for one of which, \( m=j,j+1,j+2,\ldots \), while for the other, \( m=-j,-j-1,-j-2,\ldots \) We see that any such irrep has either a lowest weight vector or a highest weight vector, but not both (like in SU(2)). Hence, the unitary irreps of the discrete series all have an infinite (but countable\footnote{This is important because a Hilbert space spanned by such a basis is then still separable. Without this property, many important results about operators on infinite dimensional Hilbert spaces would not apply.}) basis.

There are, however, two series of representations for which $j$ takes continuous values. First, there is the \emph{principal series}. For these irreps, we have
\begin{equation}
j=-\einhalb-is,~s\in\IR^+_0\komma
\end{equation}
so $j$ takes complex values with real part $-\einhalb$ and positive imaginary part. Once again, there are two irreps for every $j$, and for one of them, \( m=0,\pm1,\pm2,\ldots \), while for the other, \( m=0,\pm\einhalb,\pm\frac{3}{2},\ldots \).

The other continuous series, called \emph{exceptional series}, extrapolates from real part \( -\einhalb \) to \( j=0 \), in the sense that
\begin{equation}
	j=-\einhalb-\sigma,~\sigma\in(0,\einhalb)\punkt
\end{equation}
For this series, there exists only one irrep per $j$, for which \( m=0,\pm1,\pm2,\ldots \). We observe that all continuous representations have neither a lowest, nor a highest weight vector, which leads to sizeable computational difficulties when dealing with them.

Finally, there are two remaining so-called \emph{singleton} representations. They have either \( j=\frac{1}{4} \) or \( j=\frac{3}{4} \) and for both of them, \( m=j,j+1,j+2,\ldots \), so there is a lowest weight vector.

\section{Resolution of the Identity}\label{sec:identity}

In this appendix, we prove formula \eqref{eq:identityResolution}, the resolution of the identity. Manipulations in the case $j=\nicefrac{1}{2}$ are formal for now and we will discuss the strategy for this case later.
\begin{alignat}{2}
	&\int_{|z_0|^2 \geq 2 \ell}\dd[2]{z_0} \int_{|z_1^2|\leq |z_0|^2}\dd[2]{z_1}\frac{(2j-1)}{2\ell\pi^2}\delta(2L(z)-2\ell)\dyad{j,z} \\
	=&\int_{|z_0|^2 \geq 2 \ell}\dd[2]{z_0} \int_{|z_1^2|\leq |z_0|^2}\dd[2]{z_1}\frac{(2j-1)}{2\ell\pi^2}\delta(|z_0|^2-|z_1|^2-2\ell) \left( 1-\frac{|z_1|^2}{|z_0|^2}\right)^{2j} \label{eq:RI2}\\ \nonumber
		& ~~~~\times ~ \sum_{m,m'=j}^{\infty} \sqrt{\frac{\Gamma(m+j)}{\Gamma(m-j+1)\Gamma(2j)}\frac{\Gamma(m'+j)}{\Gamma(m'-j+1)\Gamma(2j)}} \left( \frac{z_1}{\bar{z}_0}\right)^{m-j} \left( \frac{\bar{z}_1}{{z}_0}\right)^{m'-j} \ket{j,m}\bra{j,m'} \\
	=&\int_{2 \ell}^\infty \dd{|z_0|^2} \int_0^{|z_0|^2}\dd{|z_1|^2} \frac{1}{(2 \pi)^2} \int_0^{2\pi} \dd{\phi_0}\int_0^{2\pi} \dd{\phi_1} \frac{(2j-1)}{2\ell}\delta(|z_0|^2-|z_1|^2-2\ell) \left( 1-\frac{|z_1|^2}{|z_0|^2}\right)^{2j}  \label{eq:RI3} \\ \nonumber
		& ~~~~\times ~ \sum_{m,m'=j}^{\infty} \sqrt{\frac{\Gamma(m+j)}{\Gamma(m-j+1)\Gamma(2j)}\frac{\Gamma(m'+j)}{\Gamma(m'-j+1)\Gamma(2j)}} \left( \frac{|z_1|}{|{z}_0|}\right)^{m+m'-2j}  e^{i \phi_0(m-m')} e^{i \phi_1(m-m')}  \\ \nonumber
		& ~~~~\times ~  \ket{j,m}\bra{j,m'} \\
	=&\int_{2 \ell}^\infty \dd{|z_0|^2} \int_0^{|z_0|^2}\dd{|z_1|^2} \frac{(2j-1)}{2\ell}\delta(|z_0|^2-|z_1|^2-2\ell) \left( 1-\frac{|z_1|^2}{|z_0|^2}\right)^{2j} \label{eq:RI4} \\ \nonumber
		& ~~~~\times ~ \sum_{m=j}^{\infty} \frac{\Gamma(m+j)}{\Gamma(m-j+1)\Gamma(2j)} \left( \frac{|z_1|}{|{z}_0|}\right)^{2m-2j}   \ket{j,m}\bra{j,m} \\
	=&\int_{2 \ell}^\infty \dd{|z_0|^2} \frac{(2j-1)}{2\ell} \left( \frac{2 \ell}{|z_0|^2} \right)^{2j}  \sum_{m=j}^{\infty} \frac{\Gamma(m+j)}{\Gamma(m-j+1)\Gamma(2j)} \left( \frac{|z_0|^2-2\ell}{|{z}_0|^2}\right)^{m-j}   \ket{j,m}\bra{j,m}  \label{eq:RI5} \\
	=& \sum_{m=j}^{\infty} \ket{j,m}\bra{j,m}  ~ \times ~  \left({2 \ell} \right)^{2j-1}  \frac{\Gamma(m+j)}{\Gamma(m-j+1)\Gamma(2j-1)} \int_{2 \ell}^\infty \dd{x}   \frac{\left(x-2\ell\right)^{m-j}}{x^{m+j}}  \label{eq:RI6}  \\
	=& \sum_{m=j}^{\infty} \ket{j,m}\bra{j,m}  \label{eq:RI7}
\end{alignat}
In \eqref{eq:RI2}, we just inserted the definition \eqref{eq:cohStates} of the coherent states and restricted the integration range to a subset where the argument of the delta-distribution can be zero. In \eqref{eq:RI3}, we introduced polar coordinates as $z_0 = |z_0| e^{i\phi_0}$ and $z_1 = |z_1| e^{i\phi_1}$. In \eqref{eq:RI4}, we performed the integral over $\phi_0$ and $\phi_1$, which enforces $m=m'$ and cancels the factor $1/(2\pi)^2$. In \eqref{eq:RI5}, we performed the integral over $|z_1|^2$, where the $\delta$-distribution leads to the identification $|z_1|^2 \rightarrow |z_0|^2-2\ell$. In \eqref{eq:RI6}, we rearranged the terms and relabelled $|z_0|^2 \rightarrow x$. The remaining integral in \eqref{eq:RI6} cancels its prefactors, leading to the desired result in \eqref{eq:RI7}. This last step requires $j>  \nicefrac{1}{2}$. An alternative derivation which uses a representation via harmonic oscillators can be found in \cite{LivineGroupTheoreticalQuantization}, see also the discussion in \cite{HanederMasterThesis}. 

Coming back to the case $j=\nicefrac{1}{2}$, it is possible that a resolution of the identity simply does not exist here for the coherent states, even though the coherent state system is still overcomplete. Seemingly \cite{KlauderCoherentStatePath}, this doesn't really pose a problem though, and one can try to ``act as if'' a resolution of the identity exists. Indeed, as we saw, the final phase space path integral \eqref{eq:phasespacePI} derived by this strategy does not exhibit any pathology for $j=\nicefrac{1}{2}$. 

More rigorously, we can proceed as follows. We redefine the \( \al{su}(1,1) \) discrete irreps to no longer be labelled by half-integer $j$, but by $j+\delta$, where $1\gg\delta>0$ plays the role of a regulator\footnote{In fact, there are representations of \( \al{su}(1,1) \) with real $j>\nicefrac{1}{2}$, see e.g. \cite{SchliemannCoherentStatesOf}. However, they are not representations of the group SU(1,1). For the computation at hand, this makes no difference.}. The basis states are then \( \ket{j+\delta,m+\delta} \), such that $\jz\ket{j+\delta,m+\delta}=(m+\delta)\ket{j+\delta}$ etc. In particular, one can verify that the lowest weight vector is still annihilated by \( \km \):
\begin{equation}
	\km\ket{j+\delta,j+\delta}\propto\sqrt{(j+\delta-j-\delta)(\ldots)}=0
\end{equation}
One can immediately see that in the limit $\delta\to0$, all matrix elements reproduce the usual \( \al{su}(1,1) \) values. Using this definition, we simply repeat the calculation of the resolution of the identity as before and note that all steps are well-defined. One can then use these modified states to derive the path integral formula and only take the limit $\delta\to0$ in physical expressions, i.e. matrix elements.

\end{appendix}


\begin{thebibliography}{10}

\bibitem{ThiemannModernCanonicalQuantum}
T.~Thiemann, {\em {Modern Canonical Quantum General Relativity}}.
\newblock Cambridge University Press, Cambridge, 2007.

\bibitem{RovelliBook2}
C.~Rovelli and F.~Vidotto, {\em {Covariant Loop Quantum Gravity: An Elementary
  Introduction to Quantum Gravity and Spinfoam Theory}}.
\newblock Cambridge University Press, 2014.

\bibitem{BarberoRealAshtekarVariables}
J.~Barbero, ``{Real Ashtekar variables for Lorentzian signature space-times},''
  {\em Phys. Rev. D} {\bf 51} (1995) 5507--5510, {\tt arXiv:gr-qc/9410014}.

\bibitem{PonzanoSemiclassicalLimitOf}
G.~Ponzano and T.~Regge, ``{Semiclassical limit of Racah coefficients},'' in
  {\em Spectrosc. Gr. Theor. methods Phys.} (F.~Bloch, ed.), pp.~1--58.
\newblock North-Holland Publ. Co., Amsterdam, 1968.

\bibitem{SinghLoopQuantumCosmologyABrief}
P.~Singh and I.~Agullo, ``{Loop Quantum Cosmology: A brief review},'' {\tt
  arXiv:1612.01236 [gr-qc]}.

\bibitem{AshtekarQuantumGeometryAndBlackHoleEntropy}
A.~Ashtekar, J.~Baez, A.~Corichi, and K.~Krasnov, ``{Quantum Geometry and Black
  Hole Entropy},'' {\em Phys. Rev. Lett.} {\bf 80} (1998) 904--907, {\tt
  arXiv:gr-qc/9710007}.

\bibitem{DomagalaBlackHoleEntropy}
M.~Domagala and J.~Lewandowski, ``{Black-hole entropy from quantum geometry},''
  {\em Class. Quantum Gravity} {\bf 21} (2004) 5233--5243, {\tt
  arXiv:gr-qc/0407051}.

\bibitem{HanSemiclassicalBehaviorOf}
M.~Han, ``{Semiclassical behavior of spinfoam amplitude with small spins and
  entanglement entropy},'' {\em Phys. Rev. D} {\bf 100} (2019) 084049, {\tt
  arXiv:1906.05536 [gr-qc]}.

\bibitem{BNI}
N.~Bodendorfer and Y.~Neiman, ``{Imaginary action, spinfoam asymptotics and the
  'transplanckian' regime of loop quantum gravity},'' {\em Class. Quantum
  Gravity} {\bf 30} (2013) 195018, {\tt arXiv:1303.4752 [gr-qc]}.

\bibitem{MarkopoulouCoarseGrainingIn}
F.~Markopoulou, ``{Coarse graining in spin foam models},'' {\em Class. Quantum
  Gravity} {\bf 20} (2003) 777--799, {\tt arXiv:gr-qc/0203036}.

\bibitem{OecklRENORMALIZATIONFORSPIN}
R.~Oeckl, ``{RENORMALIZATION FOR SPIN FOAM MODELS OF QUANTUM GRAVITY},'' in
  {\em Tenth Marcel Grossmann Meet.}, pp.~2296--2300, World Scientific
  Publishing Company2006.
\newblock {\tt arXiv:gr-qc/0401087}.

\bibitem{LivineCouplingOfSpacetime}
E.~R. Livine and D.~Oriti, ``{Coupling of spacetime atoms in 4D spin foam
  models from group field theory},'' {\em J. High Energy Phys.} {\bf 2007}
  (2007) 92, {\tt arXiv:gr-qc/0512002}.

\bibitem{DittrichCoarseGrainingMethods}
B.~Dittrich, F.~C. Eckert, and M.~Martin-Benito, ``{Coarse graining methods for
  spin net and spin foam models},'' {\em New J. Phys.} {\bf 14} (2012) 035008,
  {\tt arXiv:1109.4927 [gr-qc]}.

\bibitem{BahrHolonomySpinFoam}
B.~Bahr, B.~Dittrich, F.~Hellmann, and W.~Kaminski, ``{Holonomy spin foam
  models: Definition and coarse graining},'' {\em Phys. Rev. D} {\bf 87} (2013)
  044048, {\tt arXiv:1208.3388 [gr-qc]}.

\bibitem{BahrOnBackgroundIndependent}
B.~Bahr, ``{On background-independent renormalization of spin foam models},''
  {\tt arXiv:1407.7746 [gr-qc]}.

\bibitem{BahrHypercuboidalRenormalizationIn}
B.~Bahr and S.~Steinhaus, ``{Hypercuboidal renormalization in spin foam quantum
  gravity},'' {\em Phys. Rev. D} {\bf 95} (2017) 126006, {\tt arXiv:1701.02311
  [gr-qc]}.

\bibitem{BahrNumericalEvidenceFor}
B.~Bahr and S.~Steinhaus, ``{Numerical evidence for a phase transition in 4d
  spin foam quantum gravity},'' {\tt arXiv:1605.07649 [gr-qc]}.

\bibitem{CarrozzaFlowingInGroup}
S.~Carrozza, ``{Flowing in Group Field Theory Space: a Review},'' {\em SIGMA}
  {\bf 12} (2016) 070, {\tt arXiv:1603.01902 [gr-qc]}.

\bibitem{BahrRenormalizationOfSymmetry}
B.~Bahr, G.~Rabuffo, and S.~Steinhaus, ``{Renormalization of symmetry
  restricted spin foam models with curvature in the asymptotic regime},'' {\em
  Phys. Rev. D} {\bf 98} (2018) 106026, {\tt arXiv:1804.00023 [gr-qc]}.

\bibitem{DittrichCoarseGrainingFlow}
B.~Dittrich, E.~Schnetter, C.~J. Seth, and S.~Steinhaus, ``{Coarse graining
  flow of spin foam intertwiners},'' {\em Phys. Rev. D} {\bf 94} (2016) 124050,
  {\tt arXiv:1609.02429 [gr-qc]}.

\bibitem{BodendorferStateRefinementsAnd}
N.~Bodendorfer, ``{State refinements and coarse graining in a full theory
  embedding of loop quantum cosmology},'' {\em Class. Quantum Gravity} {\bf 34}
  (2017) 135016, {\tt arXiv:1607.06227 [gr-qc]}.

\bibitem{BodendorferCoarseGrainingAs}
N.~Bodendorfer and F.~Haneder, ``{Coarse graining as a representation
  change},'' {\em Phys. Lett. B} {\bf 792} (2019) 69--73, {\tt arXiv:1811.02792
  [gr-qc]}.

\bibitem{ThiemannRenormalisationReview}
T.~Thiemann, ``{Canonical Quantum Gravity, Constructive QFT and
  Renormalisation},'' {\tt arXiv:2003.13622 [gr-qc]}.

\bibitem{BojowaldDynamicalCoherentStates}
M.~Bojowald, ``{Dynamical coherent states and physical solutions of quantum
  cosmological bounces},'' {\em Phys. Rev. D} {\bf 75} (2007) 123512, {\tt
  arXiv:gr-qc/0703144}.

\bibitem{BojraDynamicsForA}
E.~F. Borja, J.~Diaz-Polo, I.~Garay, and E.~R. Livine, ``{Dynamics for a
  2-vertex quantum gravity model},'' {\em Class. Quantum Gravity} {\bf 27}
  (2010) 235010, {\tt arXiv:1006.2451 [gr-qc]}.

\bibitem{LivineGroupTheoreticalQuantization}
E.~R. Livine and M.~Martin-Benito, ``{Group theoretical quantization of
  isotropic loop cosmology},'' {\em Phys. Rev. D} {\bf 85} (2012) 124052, {\tt
  arXiv:1204.0539 [gr-qc]}.

\bibitem{BenAchourThiemannComplexifierIn}
J.~{Ben Achour} and E.~R. Livine, ``{Thiemann complexifier in classical and
  quantum FLRW cosmology},'' {\em Phys. Rev. D} {\bf 96} (2017) 066025, {\tt
  arXiv:1705.03772 [gr-qc]}.

\bibitem{BWI}
N.~Bodendorfer and D.~Wuhrer, ``{Renormalisation with SU(1, 1) coherent states
  on the LQC Hilbert space},'' {\em Class. Quantum Gravity} (2020) {\tt
  arXiv:1904.13269 [gr-qc]}.

\bibitem{BHAddendum}
N.~Bodendorfer and F.~Haneder, ``{A note on coarse graining and group
  representations}, {\tt arXiv:2011.12749 [gr-qc]}.

\bibitem{BIII}
N.~Bodendorfer, ``{Quantum reduction to Bianchi I models in loop quantum
  gravity},'' {\em Phys. Rev. D} {\bf 91} (2015) 081502(R), {\tt
  arXiv:1410.5608 [gr-qc]}.

\bibitem{BVI}
N.~Bodendorfer, ``{An embedding of loop quantum cosmology in (b,v) variables
  into a full theory context},'' {\em Class. Quantum Gravity} {\bf 33} (2016)
  125014, {\tt arXiv:1512.00713 [gr-qc]}.

\bibitem{AshtekarCastingLoopQuantum}
A.~Ashtekar, M.~Campiglia, and A.~Henderson, ``{Casting loop quantum cosmology
  in the spin foam paradigm},'' {\em Class. Quantum Gravity} {\bf 27} (2010)
  135020, {\tt arXiv:1001.5147 [gr-qc]}.
  
\bibitem{spinfoamcosmology}
Eugenio Bianchi, Carlo Rovelli, Francesca Vidotto, ``Towards Spinfoam Cosmology'', Phys. Rev. D82: 084035, 2010

\bibitem{HanSpinfoamsNearA}
M.~Han and M.~Zhang, ``{Spinfoams near a classical curvature singularity},''
  {\em Phys. Rev. D} {\bf 94} (2016) 104075, {\tt arXiv:1606.02826 [gr-qc]}.

\bibitem{HanEffectiveDynamicsFrom}
M.~Han and H.~Liu, ``{Effective dynamics from coherent state path integral of
  full loop quantum gravity},'' {\em Phys. Rev. D} {\bf 101} (2020) 046003,
  {\tt arXiv:1910.03763 [gr-qc]}.

\bibitem{BrownDustAsStandard}
J.~Brown and K.~Kuchar, ``{Dust as a standard of space and time in canonical
  quantum gravity},'' {\em Phys. Rev. D} {\bf 51} (1995) 5600--5629, {\tt
  arXiv:gr-qc/9409001}.

\bibitem{SwiezewskiOnTheProperties}
J.~Swiezewski, ``{On the properties of the irrotational dust model},'' {\em
  Class. Quantum Gravity} {\bf 30} (2013) 237001, {\tt arXiv:1307.4687
  [gr-qc]}.

\bibitem{AshtekarLoopQuantumCosmology}
A.~Ashtekar and P.~Singh, ``{Loop quantum cosmology: a status report},'' {\em
  Class. Quantum Gravity} {\bf 28} (2011) 213001, {\tt arXiv:1108.0893
  [gr-qc]}.

\bibitem{ThiemannQSD1}
T.~Thiemann, ``{Quantum spin dynamics (QSD)},'' {\em Class. Quantum Gravity}
  {\bf 15} (1998) 839--873, {\tt arXiv:gr-qc/9606089}.

\bibitem{AshtekarQuantumNatureOf}
A.~Ashtekar, T.~Pawlowski, and P.~Singh, ``{Quantum nature of the big bang:
  Improved dynamics},'' {\em Phys. Rev. D} {\bf 74} (2006) 084003, {\tt
  arXiv:gr-qc/0607039}.

\bibitem{PerelomovCoherentStatesFor}
A.~M. Perelomov, ``{Coherent states for arbitrary Lie group},'' {\em Commun.
  Math. Phys.} {\bf 26} (1972) 222--236, {\tt arXiv:math-ph/0203002}.

\bibitem{PerelomovBook}
A.~M. Perelomov, {\em {Generalized Coherent States and Their Applications}}.
\newblock Springer, Berlin, 1986.

\bibitem{AshtekarQuantumNatureOfAnalytical}
A.~Ashtekar, T.~Pawlowski, and P.~Singh, ``{Quantum nature of the big bang: An
  analytical and numerical investigation},'' {\em Phys. Rev. D} {\bf 73} (2006)
  124038, {\tt arXiv:gr-qc/0604013}.

\bibitem{SchulmanTechniquesAndApplications}
L.~S. Schulman, {\em {Techniques and applications of path integration}}.
\newblock Dover, Mineola, N.Y., 2005.

\bibitem{BergeronCoherentStatePath}
M.~Bergeron, ``{Coherent State Path Integral for the Harmonic Oscillator and a
  Spin Particle in a Constant Magnetic field},'' {\em Fortschritte der Phys.
  Phys.} {\bf 40} (1992), no.~2 119--159.

\bibitem{WilsonBreakdownOfThe}
J.~H. Wilson and V.~Galitski, ``{Breakdown of the Coherent State Path Integral:
  Two Simple Examples},'' {\em Phys. Rev. Lett.} {\bf 106} (2011) 110401, {\tt
  arXiv:1012.1328 [quant-ph]}.

\bibitem{AtlandBook}
A.~Altland and B.~Simons, {\em {Condensed matter field theory}}.
\newblock Cambridge University Press, 2010.

\bibitem{HanederMasterThesis}
F.~Haneder, {\em {Coarse Graining in Loop Quantum Cosmology}}.
\newblock Master thesis, Regensburg,
\newblock 2020.

\bibitem{BruckmannRigorousConstructionOf}
F.~Bruckmann and J.~D. Urbina, ``{Rigorous construction of coherent state path
  integrals through dualization},'' {\tt arXiv:1807.10462 [quant-ph]}.

\bibitem{ScullyBook}
M.~O. Scully and M.~S. Zubairy, {\em {Quantum optics}}.
\newblock Cambridge University Press, 2002.

\bibitem{PerezTheSpinFoam}
A.~Perez, ``{The Spin-Foam Approach to Quantum Gravity},'' {\em Living Rev.
  Relativ.} {\bf 16} (2013) {\tt arXiv:1205.2019 [gr-qc]}.

\bibitem{SteinhausCoarseGrainingSpin}
S.~Steinhaus, ``{Coarse graining spin foam quantum gravity -- a review},'' {\em
  Front. Phys.} {\bf 8} (2020) {\tt arXiv:2007.01315 [gr-qc]}.

\bibitem{RamondBook}
P.~Ramond, {\em {Group Theory. A Physicist's Survey.}}
\newblock Cambridge University Press, 2010.

\bibitem{KlauderCoherentStatePath}
J.~R. Klauder, ``{Coherent State Path Integrals Without Resolutions of
  Unity},'' {\em Found. Phys.} {\bf 31} (2001) 57--67, {\tt
  arXiv:quant-ph/0008132}.

\bibitem{SchliemannCoherentStatesOf}
J.~Schliemann, ``{Coherent states of su(1,1): correlations, fluctuations, and
  the pseudoharmonic oscillator},'' {\em J. Phys. A Math. Theor.} {\bf 49}
  (2016) 135303, {\tt arXiv:1508.04549 [quant-ph]}.




\end{thebibliography}
\end{document}